\documentclass[oneside,11pt]{article}

\usepackage[small,it]{caption}
\usepackage{url,mathrsfs,algorithm}
\usepackage{amsmath,wrapfig,color}
\usepackage{amsfonts}
\usepackage{graphicx}
\usepackage{subfigure}
\newtheorem{theorem}{Theorem}
\newtheorem{lemma}{Lemma}

\begin{document}

\title{\Huge One Scan 1-Bit Compressed Sensing}

\author{ \textbf{Ping Li} \\
         Department of Statistics and Biostatistics\\
         Department of Computer Science\\
       Rutgers University\\
         Piscataway, NJ 08854, USA\\
      \texttt{pingli@stat.rutgers.edu}
        }
\date{}
\maketitle

\begin{abstract}
\noindent Based
on $\alpha$-stable random projections with small $\alpha$, we develop a simple algorithm for compressed sensing (sparse signal recovery) by utilizing only the signs (i.e., 1-bit) of the measurements.  Using only 1-bit information of the measurements  results in substantial cost reduction  in collection, storage, communication, and decoding for compressed sensing.  The proposed algorithm is efficient in that the decoding procedure requires only one scan of the coordinates. Our analysis can precisely show that, for a $K$-sparse signal of length $N$, $12.3K\log N/\delta$ measurements (where $\delta$ is the confidence) would be sufficient for recovering the support and the signs of the signal. While the method is very robust against typical measurement noises, we also provide the analysis of the scheme under   random flipping of the signs of the measurements. \\

\noindent Compared to the well-known work on 1-bit marginal regression (which can also be viewed as a one-scan method), the proposed algorithm requires orders of magnitude fewer measurements. Compared to  1-bit Iterative Hard Thresholding (IHT) (which is not a one-scan algorithm), our method is still significantly more accurate. Furthermore, the proposed method is reasonably robust against random sign flipping while IHT is known to be very sensitive to this type of noise.

\end{abstract}

\section{Introduction}\label{sec:intro}

Compressed sensing (CS)~\cite{Article:Donoho_CS_JIT06,Article:Candes_Robust_JIT06} is a popular and important topic in mathematics and engineering, for recovering sparse signals from  linear measurements. Here, we consider a $K$-sparse signal of length $N$, denoted by $x_i$, $i = 1$ to $N$. In our scheme, the linear measurements are collected as follows
\begin{align}\notag
y_j = \sum_{i=1}^N x_i s_{ij}, \ \ j = 1, 2, ..., M,  \hspace{0.1in} \text{ where } s_{ij} \sim S(\alpha,1)
\end{align}
where $y_j$'s are the measurements and $s_{ij}$ is the $(i,j)$-th entry of the design matrix  sampled i.i.d. from an $\alpha$-stable distribution with unit scale, denoted by $S(\alpha,1)$.  This is different from classical framework of compressed sensing.   Classical algorithms of compressed sensing use  Gaussian design (i.e., $\alpha=2$ in the family of stable distribution) or Gaussian-like design (e.g., a distribution with finite variance), to recover signals  via computationally intensive methods such as linear programming~\cite{Article:Chen98} or greedy  methods such as {orthogonal matching pursuit (OMP)}~\cite{Proc:Pati93,Article:Mallat93,Article:CoSaMP_09,Article:Zhang_RIP11}.

\newpage

The recent work~\cite{Proc:CCCS_COLT14} studied the use of $\alpha$-stable random projections with $\alpha<2$, for accurate one-scan compressed sensing. Basically, if $Z\sim S(\alpha,1)$, then its characteristic function is $E\left(e^{\sqrt{-1}Zt}\right) = e^{-|t|^\alpha}$, where $0<\alpha\leq2$. Thus, both Gaussian ($\alpha=2$) and Cauchy ($\alpha=1$) distributions are special instances of the $\alpha$-stable distribution family. Inspired by~\cite{Proc:CCCS_COLT14}, we develop \textbf{one scan 1-bit compressed sensing} by using small $\alpha$ (e.g., $\alpha=0.05$) and only the sign information (i.e., $sgn(y_j)$) of the measurements. Compared to alternatives, the proposed method is  fast and accurate.\\

The problem of 1-bit compressed sensing has been studied in the literature of statistics, information theory and machine learning, e.g., \cite{Proc:Boufounos08,Article:1Bit_IT13,Proc:1BitCS_ICML13,Article:Plan_IT13,Proc:Chen_AISTATS15,Proc:Slawski_NIPS15}. 1-bit compressed sensing has many advantages. When the measurements are collected, the hardware will anyway have to quantize the measurements. Also, using only the signs will potentially reduce the cost of storage and transmission (if the number of measurements does not have to increase too much). It appears, however, that the current methods for 1-bit compressed sensing have not  fully accomplished those goals. For example, \cite{Article:1Bit_IT13} showed that even with $M/N=2$ (i.e., the number of measurements is twice as the length of signal), there are still noticeable recovery errors in their experiments. A  recent work~\cite{Proc:Chen_AISTATS15} also reported that even when the number of measurements exceeds length of the signal, the errors are still observable.

In the experimental study in Section~\ref{sec_exp}, our comparisons with 1-bit marginal regression~\cite{Article:Plan_IT13,Proc:Slawski_NIPS15} illustrate that the proposed method needs orders of magnitude fewer measurements. Compared to 1-bit Iterative Hard Thresholding (IHT)~\cite{Article:1Bit_IT13}, our algorithm is still significantly more accurate. Furthermore, while our method is reasonably robust against random sign flipping, IHT is known to be very sensitive to that kind of noise. \\

A distinct  advantage  of our proposed method is that, largely due to the one-scan nature,  we can very precisely analyze the  algorithm with or without random flipping noise; we also provide the precise constants of the bounds. For example,  even for a conservative version of our algorithm,  the required number of measurements, with probability $>1-\delta$, would be no more than $12.3K\log N/\delta$ (and the practical performance is even  better). Here $\delta$ (e.g., 0.05) is the  notation for confidence.\\

The method of Gaussian (i.e., $\alpha=2$) random projections has become extremely popular in  machine learning and information theory (e.g.,~\cite{Proc:Frund_NIPS08}). The use of $\alpha$-stable random projections was previously studied in the context of estimating the $l_\alpha$ norms  (e.g., $\sum_{i=1}^N|x_i|^\alpha$) of \textbf{data streams}, in the theory literature~\cite{Article:Indyk_JACM06,Proc:Li_SODA08} as well as in machine learning venue~\cite{Proc:Li_Hastie_NIPS07}. Consequently, our 1-bit CS algorithm also inherits the advantage when the data  (signals) arrive in a streaming fashion~\cite{Article:Muthukrishnan_05}.\\

The recent work~\cite{Proc:CCCS_COLT14} used $\alpha$-stable projections with very small $\alpha$ to recover sparse signals, with many significant advantages: (i) the algorithm needs only one scan; (ii) the method is extremely robust against measurement noises (due to the heavy-tailed nature of the projections); and (iii) the recovery procedure is per coordinate in that even when there are no sufficient measurements, a significant portion of the nonzero coordinates can still be recovered. The major disadvantage of~\cite{Proc:CCCS_COLT14}  is that, since the measurements are also heavy-tailed, the required storage for the measurements  might be substantial. Our proposed 1-bit algorithm provides one practical (and  simple) solution.

\section{The  Proposed Algorithm}

In our algorithm, the entries (i.e., $s_{ij}$) of the design matrix are sampled from i.i.d.  $\alpha$-stable with unit scale, denoted by $S(\alpha,1)$. We can follow the classical procedure to generate samples~\cite{Article:Chambers_JASA76} from $S(\alpha,1)$. That is, we first sample independent exponential  $w\sim exp(1)$ and uniform $u\sim unif(-\pi/2,\pi/2)$  variables, then
\begin{align}\label{eqn_stable_sample}
g(u,w;\alpha) = \frac{\sin(\alpha u)}{(\cos u)^{1/\alpha}}
\Big[\frac{\cos(u-\alpha u)}{w}\Big]^{(1-\alpha)/\alpha}\sim S(\alpha,1)
\end{align}
There are two excellent books on stable distributions~\cite{Book:Zolotarev_86,Book:Samorodnitsky_94}. Basically, if $Z\sim S(\alpha,1)$, then its characteristic function is $E\left(e^{\sqrt{-1}Zt}\right) = e^{-|t|^\alpha}$. However, closed-form expressions of the density exists only for $\alpha=2$ (i.e., Gaussian), $\alpha=1$ (i.e., Cauchy), or $\alpha=0+$.\\

 Alg.~\ref{alg_proposed} summarizes our  one-scan algorithm for recovering the signs of sparse signals.
\begin{algorithm}{
\textbf{Input:} $K$-sparse signal $\mathbf{x}\in\mathbb{R}^{1\times N}$, design matrix $\mathbf{S}\in\mathbb{R}^{N\times M}$ with entries sampled from $S(\alpha,1)$ with small $\alpha$ (e.g., $\alpha=0.05$). To generate the $(i,j)$-th entry $s_{ij}$,  we sample  $u_{ij} \sim uniform(-\pi/2,\ \pi/2)$ and  $w_{ij}\sim exp(1)$ and  compute $s_{ij} = g(u_{ij},w_{ij};\alpha)$ by (\ref{eqn_stable_sample}).

\vspace{0.1in}

\textbf{Collect:} Linear measurements:\  $y_j = \sum_{i=1}^N x_i s_{ij}$, $j = 1$ to $M$.

\vspace{0.1in}

\textbf{Compute:}  For each coordinate $i =1$ to $N$, compute
{\small\begin{align}\notag
&Q_i^{+} = \sum_{j=1}^M\log \left(1+{sgn(y_j)}{sgn(u_{ij})}e^{-\left({K}-1\right)w_{ij}}\right),\\\notag
&Q_i^{-}  = \sum_{j=1}^M\log \left(1-{sgn(y_j)}{sgn(u_{ij})}e^{-\left({K}-1\right)w_{ij}}\right)
\end{align} }

\textbf{Output:} For  $i=1$ to $N$,  report the estimated sign:
$\hat{sgn(x_i)} = \left\{
\begin{array}{ll}
+ 1 & \text{if } Q_i^+>0 \\
- 1 & \text{if } Q_i^->0\\
0 & \text{if } Q_i^+ <0 \text{ and } Q_i^-<0
\end{array}
\right.$

}\caption{Stable measurement collection and the one scan 1-bit algorithm for sign recovery. }
\label{alg_proposed}
\end{algorithm}

The central component of the algorithm is to compute $Q_i^+$ and $Q_i^-$, for $i=1$ to $N$, where
\begin{align}\label{eqn_Q+}
&Q_i^{+} = \sum_{j=1}^M\log \left(1+{sgn(y_j)}{sgn(u_{ij})}e^{-\left({K}-1\right)w_{ij}}\right)\\\label{eqn_Q-}
&Q_i^{-}  = \sum_{j=1}^M\log \left(1-{sgn(y_j)}{sgn(u_{ij})}e^{-\left({K}-1\right)w_{ij}}\right)
\end{align}
Later we will explain that it makes no essential difference   if we replace $sgn(u_{ij})$ with $sgn(s_{ij})$ and $w_{ij}$ with $1/|s_{ij}|^\alpha$.  The parameter $\alpha$ should be  reasonably small, e.g., $\alpha = 0.05$.  In many prior studies of compressed sensing, $K$ is often assumed to be known. Very interestingly, even if $K$ is unknown, it can still be  reliably estimated in our framework using only a very small number (e.g., 5) of measurements, as validated  in Sec.~\ref{sec_K_est}.

To make the theoretical analysis easier,  Alg.~\ref{alg_proposed} uses ``0'' as the threshold for estimating the sign:
\begin{align}\label{eqn_RecoveryQ0}
\hat{sgn(x_i)} = \left\{
\begin{array}{ll}
+ 1 & \text{if } Q_i^+>0 \\
- 1 & \text{if } Q_i^->0\\
0 & \text{if } Q_i^+ <0 \text{ and } Q_i^-<0
\end{array}
\right.
\end{align}
Later in the paper, Lemma~\ref{lem_Q} will show that at most one of $Q_i^+$ and $Q_i^-$ can be positive. Using 0 as the threshold  simplifies the analysis. As will be shown in our experiments, a more practical version of the algorithm will  reduce the number of measurements predicted by the  analysis.\\

Note that, unless  the signal is  ternary (i.e., $x_i\in\{-1,0,1\}$), we will need  another  procedure for estimating the values of the nonzero entries. A simple strategy is to do a least square  on the reported coordinates, by collecting $K$ additional measurements. \\

Next, we will  present the intuition and theory for the proposed algorithm.

\section{Intuition}

Our proposed algorithm, through the use of $Q_i^+$ and $Q_i^-$, is based on the joint likelihood of $(sgn(y_j),s_{ij})$. Denote the density function of $S(\alpha,1)$ by $f_S(s)$. Recall
\begin{align}
y_j = \sum_{t=1}^N x_t s_{tj} = x_i s_{ij} + \sum_{t\neq i} x_t s_{tj} = x_is_{ij} + \theta_iS_j
\end{align}
where $S_j\sim S(\alpha,1)$ is independent of $s_{ij}$ and  $\theta_i = \left(\sum_{t\neq i }|x_t|^\alpha\right)^{1/\alpha}$. Using a  conditional probability argument, the joint density of $(y_j,s_{ij})$ can be shown to be $\frac{1}{\theta_i}f_S(s_{ij})f_S\left(\frac{y_j-x_is_{ij}}{\theta_i}\right)$. Now, suppose we only use (store) the sign information of $y_j$. We have
\begin{align}\notag
\mathbf{Pr}\left(y_j>0, s_{ij}\right) =& \int_0^\infty \frac{1}{\theta_i}f_S(s_{ij})f_S\left(\frac{y-x_is_{ij}}{\theta_i}\right) dy\\\notag
 =&f_S(s_{ij})\left(1-F_S\left(\frac{-x_is_{ij}}{\theta_i}\right)\right)\\\notag
 =&f_S(s_{ij}) F_S\left(\frac{x_is_{ij}}{\theta_i}\right)
\end{align}
where $F_S$ is the cumulative distribution function (cdf) of $S(\alpha,1)$.  Similarly,
\begin{align}\notag
\mathbf{Pr}\left(y_j<0, s_{ij}\right) =& \int_{-\infty}^0 \frac{1}{\theta_i}f_S(s_{ij})f_S\left(\frac{y-x_is_{ij}}{\theta_i}\right) dy\\\notag
 =&  f_S(s_{ij}) F_S\left(-\frac{x_is_{ij}}{\theta_i}\right)
\end{align}
which means the joint log-likelihood is proportional  to
$l(x_i,\theta_i) =\sum_{j=1}^M \log F_S\left(sgn(y_j)\frac{x_is_{ij}}{\theta_i}\right)$.\\

Since our algorithm uses small $\alpha$, we can take advantage of the limit density at $\alpha=0+$. Suppose $u\sim uniform(-\pi/2,\pi/2)$ and $w\sim exp(1)$. Using (\ref{eqn_stable_sample}), we can express $Z=g(u,w;\alpha)\approx sgn(u)/w^{1/\alpha}$. In other words, in the limit $\alpha\rightarrow0+$, $1/|Z|^\alpha \sim exp(1)$. This fact was originally established by~\cite{Article:Cressie_75} and was used by~\cite{Proc:Li_SODA08} to derive the harmonic mean estimator (\ref{eqn_Khat}) of $K$.\\

Therefore, as $\alpha\rightarrow0+$, we can write the cdf as  $F_S(s)= \frac{1}{2}+sgn(s)\frac{1}{2}e^{-|s|^{-\alpha}}$, which leads to
\begin{align}\notag
l(x_i,\theta_i) = \sum_{j=1}^M\log \left(1+sgn(s_{ij}x_iy_j)\exp\left(-\left|\frac{\theta_i}{x_is_{ij}}\right|^{\alpha}\right)
\right)
\end{align}
Clearly, if $x_i=0$, then $l(x_i,\theta_i) = 0$. This is the reason why it is convenient to use 0 as the threshold. We can then use the following $Q_i^+$ and $Q_i^-$  to determine if $x_i>0$ or $x_i<0$:
{\small\begin{align}\notag
&Q_i^{+} = \sum_{j=1}^M\log \left(1+sgn(s_{ij}y_j)\exp\left(-\frac{K-1}{|s_{ij}|^\alpha}\right)\right),\\\notag
&Q_i^{-} = \sum_{j=1}^M\log \left(1-sgn(s_{ij}y_j)\exp\left(-\frac{K-1}{|s_{ij}|^\alpha}\right)\right)
\end{align}}\vspace{-0.15in}

As $\alpha\rightarrow0+$, we have $\theta_i^\alpha=K-1$ (if $x_i\neq 0$) or $K$ (if $x_i=0$). Also note that $|x_i|^\alpha = 0$ (if $x_i=0$) or 1 (if $x_i\neq 0$).  Because $sgn(s_{ij}) = sgn(u_{ij})$ and $\frac{1}{|s_{ij}|^\alpha}$ becomes $w_{ij}$, we can write them as
\begin{align}\notag
&Q_i^{+} = \sum_{j=1}^M\log \left(1+{sgn(y_j)}{sgn(u_{ij})}e^{-\left({K}-1\right)w_{ij}}\right),\\\notag
&Q_i^{-}  = \sum_{j=1}^M\log \left(1-{sgn(y_j)}{sgn(u_{ij})}e^{-\left({K}-1\right)w_{ij}}\right)
\end{align}
This is  the reason why we compute  $Q_i^{+}$ and $Q_i^{-}$  as in (\ref{eqn_Q+}) and (\ref{eqn_Q-}), respectively.\\

So far, we have explained the idea behind our  proposed Alg.~\ref{alg_proposed}. Next we will conduct further theoretical analysis for the error probabilities and consequently the sample complexity bound.

\section{Analysis}

Our analysis  will  repeatedly use the fact that
$sgn(s_{ij} y_j) = sgn(y_j/s_{ij}) = sgn(x_i + \theta_i S_j/s_{ij})$, where $S_j\sim S(\alpha,1)$ is independent of $s_{ij}$ and  $\theta_i = \left(\sum_{t\neq i }|x_t|^\alpha\right)^{1/\alpha}$.  Note that both $s_{ij}$ and $y_j$ are symmetric random variables.

Our first lemma says that at most one of  $Q_i^+$ and $Q_i^-$, respectively defined in  (\ref{eqn_Q+}) and (\ref{eqn_Q-}), can be positive.

\begin{lemma}\label{lem_Q}
If $Q_{i}^+>0$ then $Q_i^-<0$. If $Q_{i}^->0$  then $Q_i^+<0$.

\noindent \textbf{Proof}:\hspace{0.1in} It is more convenient to examine $e^{Q_i^+}$ and $e^{Q_i^-}$ and compare them with 1. Let $z_j=e^{-(K-1)w_{ij}}$. Note that  $0<z_j<1$.
Now suppose $e^{Q_i^+}>1$. We divide the coordinates, $j=1$ to $M$, into two disjoint sets $I$ and $II$, such that
\begin{align}\notag
e^{Q_i^+}=\prod_{j\in I}|1+z_j| \prod_{j\in II}|1-z_j|>1
\end{align}
Because $\frac{1}{1-z_j} >1+z_j$ and $\frac{1}{1+z_j} > 1-z_j$, we must have
\begin{align}\notag
\prod_{j\in I}\left|\frac{1}{1-z_j}\right| \prod_{j\in II}\left|\frac{1}{1+z_j}\right|> \prod_{j\in I}|1+z_j| \prod_{j\in II}|1-z_j|>1
\end{align}
which means we must have
\begin{align}\notag
e^{Q_i^-}=  \prod_{j\in I}\left|{1-z_j}\right| \prod_{j\in II}\left|{1+z_j}\right|<1
\end{align} This completes the proof.$\hfill\Box$
\end{lemma}

Although Lemma~\ref{lem_Q} suggests that it is convenient to use 0 as the threshold, we provide more general error probability tail bounds by comparing $Q_i^+$ and $Q_i^-$ with $\epsilon M/K$, where $\epsilon$ is not necessarily nonnegative. The following intuition might be helpful to  see why $M/K$ is the right scale:
\begin{align}\notag
|Q_i^+| =& \left|\sum_{j=1}^M\log \left(1+sgn(y_j/s_{ij})\exp\left(-(K-1)w_{ij}\right)\right)\right| \\\notag
\leq&\sum_{j=1}^M\left|\log \left(1+sgn(y_j/s_{ij})\exp\left(-(K-1)w_{ij}\right)\right)\right|\\\notag
\approx& \sum_{j=1}^M \exp\left(-(K-1)w_{ij}\right)
\end{align}
By the moment  generating function of exponential distribution, we know that
\begin{align}\notag
&E\left(\sum_{j=1}^M \exp\left(-(K-1)w_{ij}\right)\right)%\\\notag
= \sum_{j=1}^M E\exp\left(-(K-1)w_{ij}\right)
= \frac{M}{(1+K-1)} = \frac{M}{K}
\end{align}\\

Lemma~\ref{lem_ProbH1} concerns the error probability (i.e., the false positive) when $x_i=0$ and $\epsilon M/K$ is used as the threshold.

\begin{lemma}\label{lem_ProbH1}
For any $\epsilon$ and any $t\geq 0$, we have
\begin{align}
&\mathbf{Pr}\left(Q_i^+>\epsilon M/K, x_i=0\right)
=\mathbf{Pr}\left(Q_i^->\epsilon M/K, x_i=0\right)\leq\exp\left\{- \frac{M}{K}H_1(t;\epsilon,K)\right\}
\end{align}
where
\begin{align}\label{eqn_H1K}
H_1(t;\epsilon,K) =& \epsilon t-K\log \left(1  + \frac{t(t-1)}{(2K-1)2!} +  \frac{t(t-1)(t-2)(t-3)}{(4K-3)4!}+...\right)\\\notag
=& \epsilon t - K\log\left(1+\sum_{n=2,4,6,...}^\infty \frac{1}{nK-n+1}\prod_{l=0}^{n-1}\frac{t-l}{n-l}\right)
\end{align}
In the limit as $K\rightarrow\infty$, we have
\begin{align}\label{eqn_H1inf}
H_1(t;\epsilon,\infty) =& \epsilon t  - \left(\frac{t(t-1)}{2\times 2!} +  \frac{t(t-1)(t-2)(t-3)}{4\times4!}+...\right)\\\notag
 =& \epsilon t - \sum_{n=2,4,6,...}^\infty \frac{1}{n}\prod_{l=0}^{n-1}\frac{t-l}{n-l}
\end{align}
\textbf{Proof:}\hspace{0.2in} See Appendix~\ref{app_lem_ProbH1}.$\hfill\Box$
\end{lemma}

\begin{figure}[h!]
\begin{center}
\mbox{
\includegraphics[width = 2.5in]{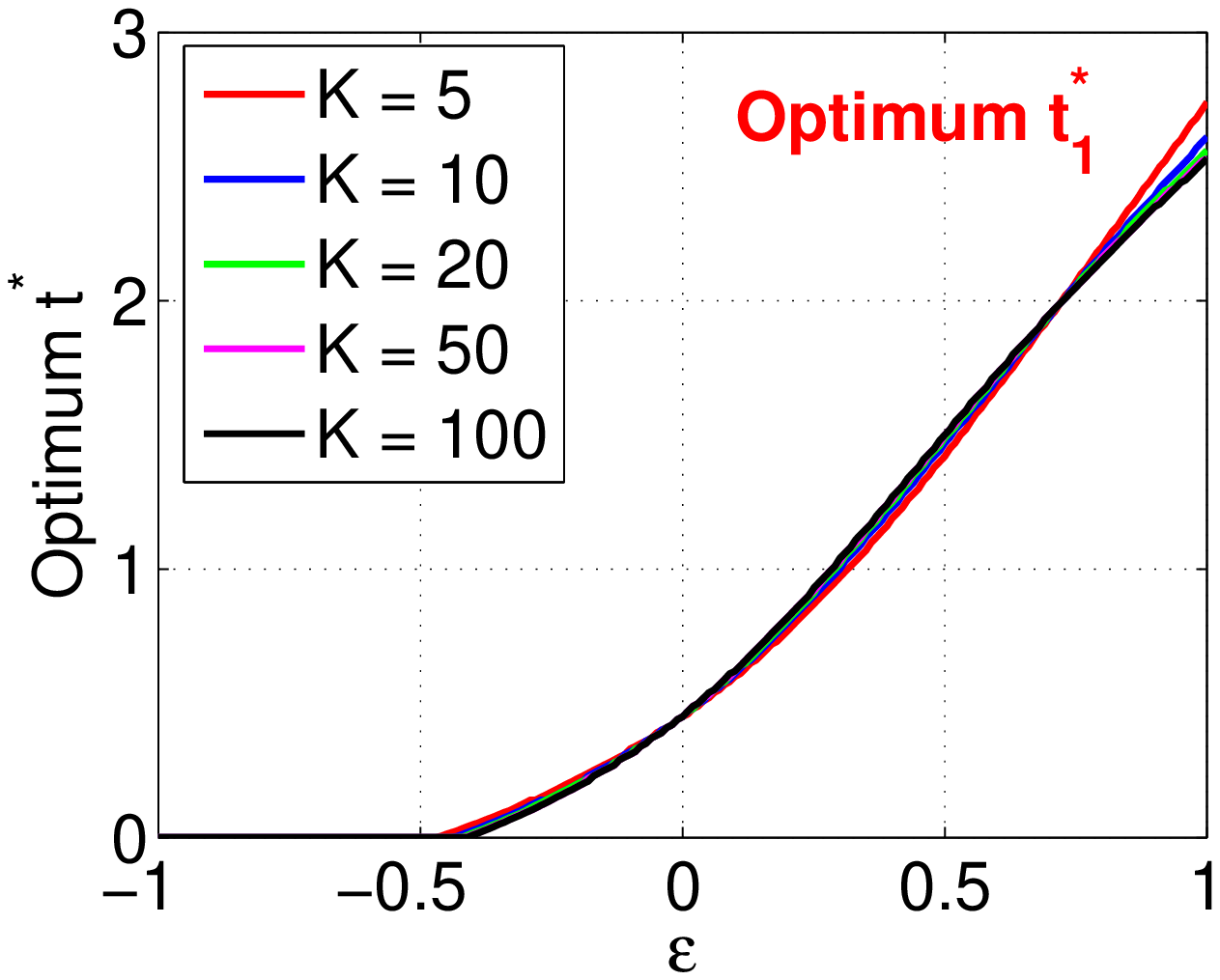}\hspace{0.3in}
\includegraphics[width = 2.5in]{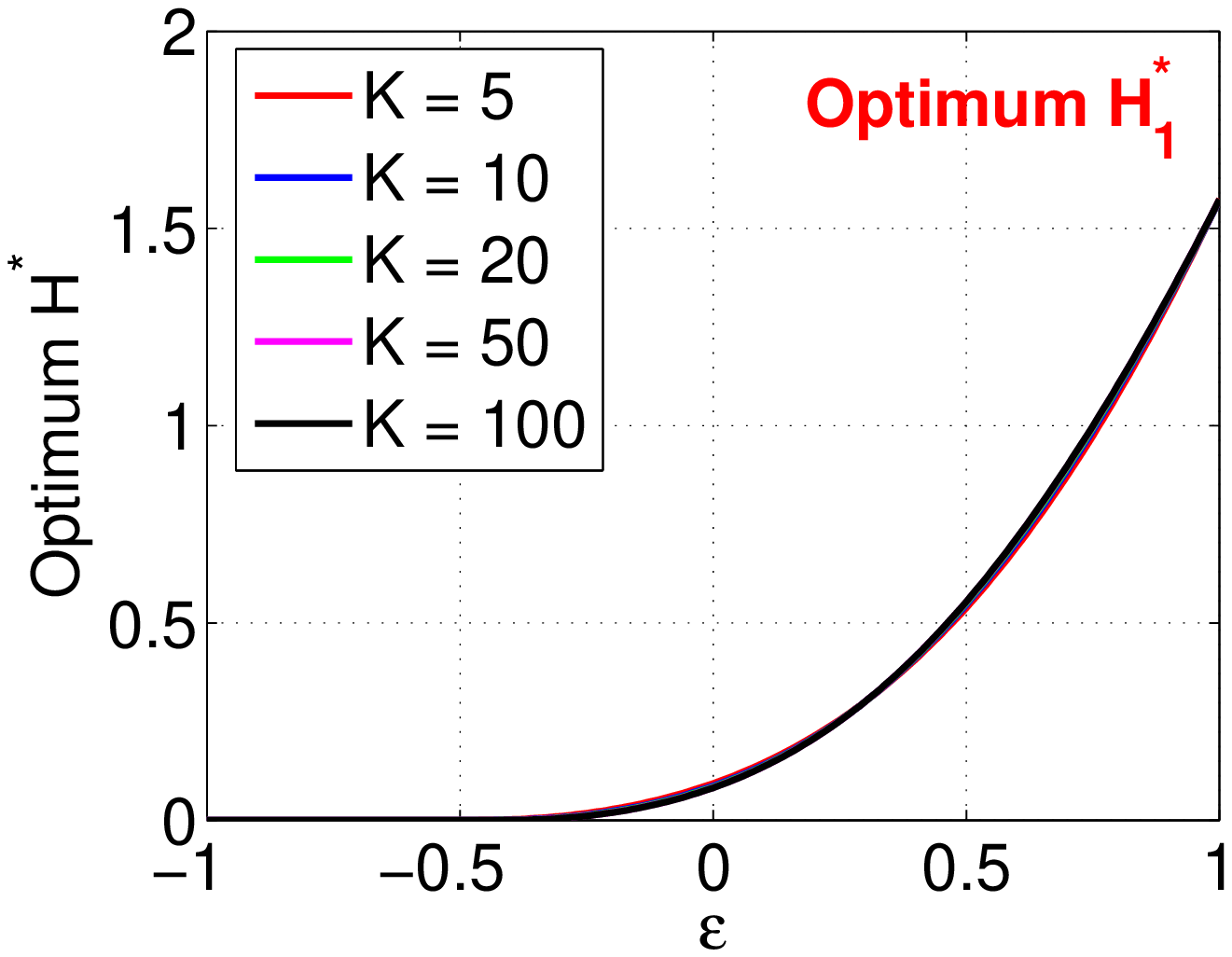}
}
\end{center}
\vspace{-0.2in}
\caption{For Lemma~\ref{lem_ProbH1}, we plot the optimum $t=t_1^*$ values (left panel) which maximizes $H_1(t;\epsilon,K)$, as well as the optimum values $H_1 = H_1^*$ at $t=t_1^*$ (right panel), for $K=5$ to 100. The different curves essentially overlap. At the threshold $\epsilon=0$, the value $1/H_1^*$ is about 12.2 (and smaller than 12.3).    }\label{fig_OptH1}
\end{figure}

To minimize the error probability in Lemma~\ref{lem_ProbH1}, we need to seek the optimum (maximum) values of $H_1$ for given $\epsilon$ and $K$. Figure~\ref{fig_OptH1} plots the optimum values $t = t_1^*$  as well as the optimum values of $H_1^*$ for $K=5$ to 100. As expected, these optimum values are insensitive to $K$ (in fact, no essential difference from the limiting case of $K\rightarrow\infty$). At $\epsilon = 0$, the value of $1/H_1^*$ is about 12.2. Note that to control the error probability to be $<\delta$, the required number of measurements will be $M\geq \frac{K}{H_1^*}\log N/\delta$. Thus we use a numerical number $12.3$ for the bound of the sample complexity.

\newpage

Next, Lemma~\ref{lem_ProbH2} concerns the false negative error probability when $x_i\neq 0$.

\begin{lemma}\label{lem_ProbH2}
For any $\epsilon$, $0<t<1$, and $\alpha\rightarrow0$,  we have
\begin{align}
\mathbf{Pr}\left(Q_i^+<\epsilon M/K, x_i>0\right)
=&\mathbf{Pr}\left(Q_i^-<\epsilon M/K, x_i<0\right)\leq \exp\left(-\frac{M}{K}H_2(t;\epsilon,K)\right)
\end{align}
where
%\begin{align}
%&H_2(t;\epsilon,K) = -\epsilon t - K\times\log \left[\begin{array}{c}
%1 + \sum_{n=2,4,6...}^\infty\frac{1}{n(K-1)+1} \prod_{l=0}^{n-1}\frac{t-l}{n-l}\\\\-
%\sum_{n=1,3,5...}^\infty\frac{1}{(n+1)(K-1)+1} \prod_{l=0}^{n-1}\frac{t-l}{n-l}
%\end{array}\right]
%\end{align}
\begin{align}
&H_2(t;\epsilon,K) = -\epsilon t - K\times \log\left[A\right]
\end{align}
\begin{align}\notag
A =& 1 + \sum_{n=2,4,6...}^\infty\frac{1}{n(K-1)+1} \prod_{l=0}^{n-1}\frac{t-l}{n-l}%\\\notag
-
\sum_{n=1,3,5...}^\infty\frac{1}{(n+1)(K-1)+1} \prod_{l=0}^{n-1}\frac{t-l}{n-l}
\end{align}
and
\begin{align}
&H_2(t;\epsilon,\infty) = -\epsilon t -
\sum_{n=2,4,6...}^\infty\frac{1}{n} \prod_{l=0}^{n-1}\frac{t-l}{n-l}+\sum_{n=1,3,5...}^\infty\frac{1}{(n+1)} \prod_{l=0}^{n-1}\frac{t-l}{n-l}
\end{align}
\textbf{Proof:}\ \ See Appendix~\ref{app_lem_ProbH2}.$\hfill\Box$
\end{lemma}

\begin{figure}[h!]
\begin{center}
\mbox{
\includegraphics[width = 2.5in]{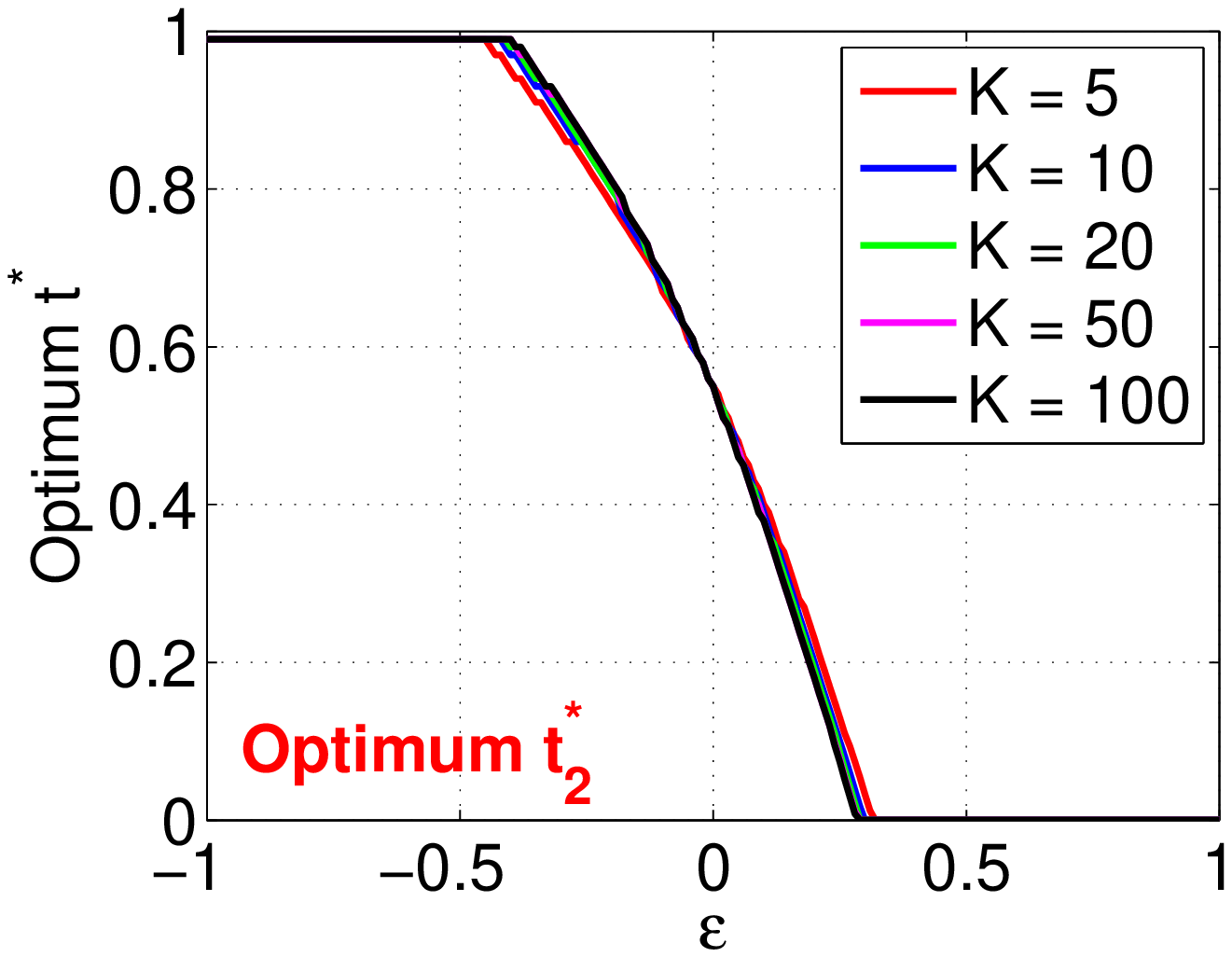}\hspace{0.3in}
\includegraphics[width = 2.5in]{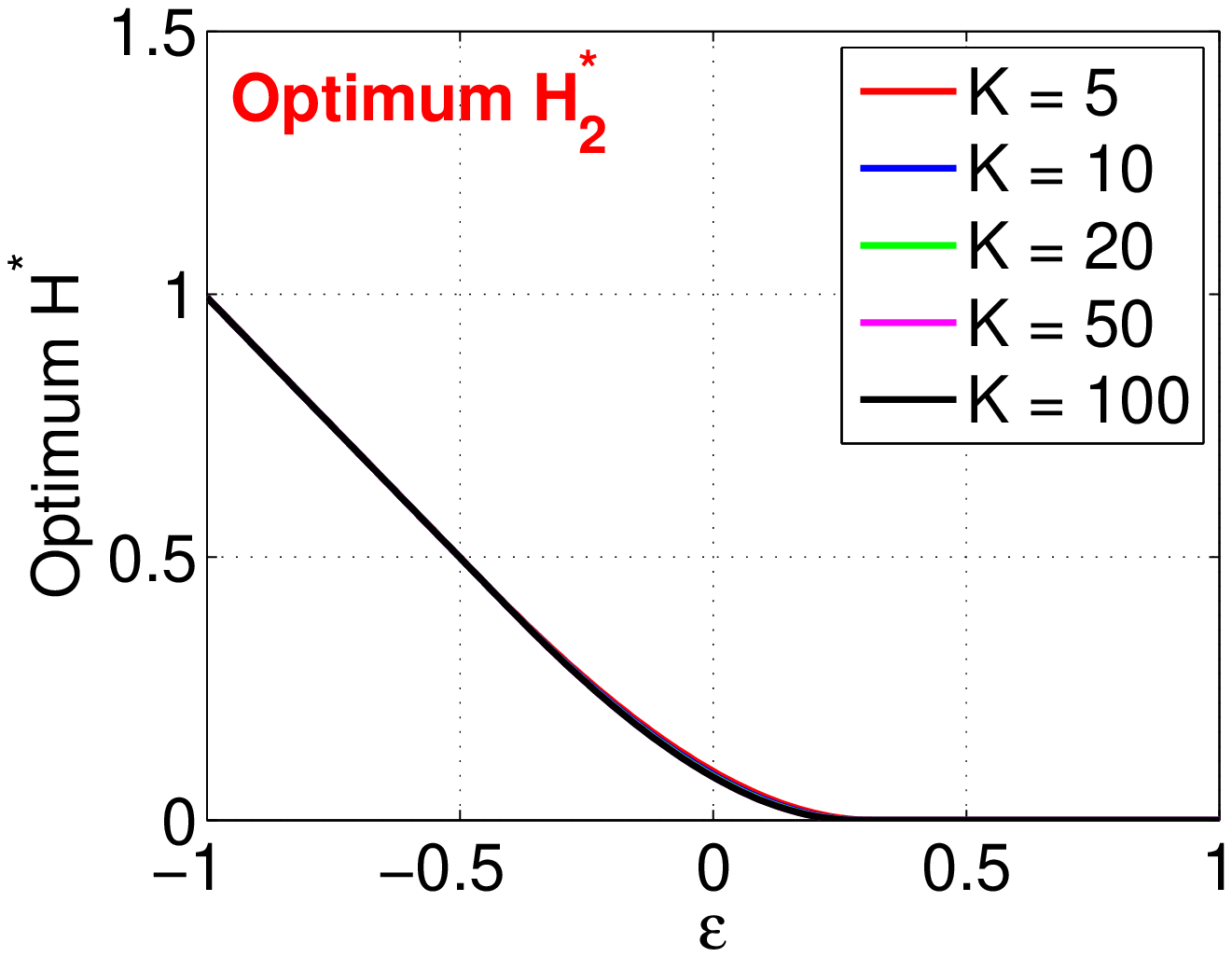}
}
\end{center}
\vspace{-0.2in}
\caption{For Lemma~\ref{lem_ProbH2}, we plot the optimum $t=t_2^*$ values (left panel) which maximizes $H_2(t;\epsilon,K)$, as well as the optimum values $H_2 = H_2^*$ at $t=t_2^*$ (right panel), for $K=5$ to 100. The different curves essentially overlap. At $\epsilon=0$, the value of $1/H_2^*$ is again about 12.2 (which is smaller than 12.3).    }\label{fig_OptH2}
\end{figure}
Figure~\ref{fig_OptH2} plots the optimum $t_2^*$ values which maximize $H_2$, together with the optimum $H_2^*$ values. Interestingly, when $\epsilon=0$, the value of $1/H_2^*$ is also about 12.2 (smaller than 12.3). This is not surprising, because, for both $H_1(t;\epsilon,\infty)$ and $H_2(t;\epsilon,\infty)$, the leading term at $\epsilon=0$ is $\frac{t(t-1)}{4}$.\\

\noindent\textbf{Sample Complexity}. Given $K$, $N$, $\epsilon$, $\delta$, the required  number measurements  can be  computed from
\begin{align}\notag
&(N-K)\times \mathbf{Pr}\left(Q_i^+>\epsilon M/K, x_i=0\right)+ K \times \mathbf{Pr}\left(Q_i^+<\epsilon M/K, x_i>0\right) \leq \delta
\end{align}
When $\epsilon=0$, because the constants of both error probabilities are upper bounded by 12.3, we obtain a convenient expression of complexity, which we present as  Theorem~\ref{thm_RecoveryQ0}.
\begin{theorem}\label{thm_RecoveryQ0}
Using Alg.~\ref{alg_proposed}, in order for the total error (for estimating the signs) of all the coordinates to be bounded by some $\delta>0$, it suffices to use $M = \lceil12.3K\log N/\delta\rceil$ measurements.

\end{theorem}

\section{Recovery Under Noise}

We can add measurement noises: $y_j = \sum_{i=1}^N x_i s_{ij} + n_j$, where typically $n_j \sim N(0,\sigma^2)$ at some noise level $\sigma$. The framework of sparse recovery using $\alpha$-stable random projections with small $\alpha$ is extremely (or boringly) robust against this type of measurement noises~\cite{Proc:CCCS_COLT14}. To make the study more interesting, we consider another common noise model for 1-bit compressed sensing by randomly flipping the signs of the measurements.\\

That is, we introduce independent variables $r_j$, $j=1$ to $M$, so that $r_j = 1$ with probability $1-\gamma$ and 0 with probability $\gamma$. During recovery, we use $(r_j y_j)$ to replace the original $y_j$.  To differentiate from the previous notation, we use $Q_{i,\gamma}^+$ and $Q_{i,\gamma}^-$, respectively, to replace $Q_{i}^+$ and $Q_{i}^-$.\\

Interestingly, Lemma~\ref{lem_ProbH3} shows that random flipping does not affect the false positive probability.
\begin{lemma}\label{lem_ProbH3}
For any $\epsilon$ and any $t\geq 0$, we have
\begin{align}
&\mathbf{Pr}\left(Q_{i,\gamma}^+>\epsilon M/K, x_i=0\right)=\mathbf{Pr}\left(Q_{i,\gamma}^->\epsilon M/K, x_i=0\right)\leq\exp\left\{- \frac{M}{K}H_1(t;\epsilon,K)\right\}
\end{align}
where $H_1(t;\epsilon,K)$ is the same as in Lemma~\ref{lem_ProbH1}.\\

\noindent\textbf{Proof:}\hspace{0.2in} See Appendix~\ref{app_lem_ProbH3}. The key is that $sgn(r_j u_{ij})$ and $sgn(u_{ij})$ has the same distribution.  $\hfill\Box$\\
\end{lemma}

On the other hand, as shown in the next lemma, this randomly flipping (with probability $\gamma$) does affect the false negative probability.
\begin{lemma}\label{lem_ProbH4}
For any $\epsilon$, $0<t<1$, and $\alpha\rightarrow0$,  we have
\begin{align}
&\mathbf{Pr}\left(Q_{i,\gamma}^+<\epsilon M/K, x_i>0\right)
=\mathbf{Pr}\left(Q_{i,\gamma}^-<\epsilon M/K, x_i<0\right)\leq \exp\left(-\frac{M}{K}H_4(t;\epsilon,K,\gamma)\right)
\end{align}
\begin{align}
&H_4(t;\epsilon,K,\gamma) = -\epsilon t - K\times\log\left[B\right]
\end{align}
\begin{align}\notag
B =& 1 + \sum_{n=2,4,6...}^\infty\frac{1}{n(K-1)+1} \prod_{l=0}^{n-1}\frac{t-l}{n-l} -
\sum_{n=1,3,5...}^\infty\frac{1-2\gamma}{(n+1)(K-1)+1} \prod_{l=0}^{n-1}\frac{t-l}{n-l}
\end{align}
\begin{align}
&H_4(t;\epsilon,\infty,\gamma) = -\epsilon t -
\sum_{n=2,4,6...}^\infty\frac{1}{n} \prod_{l=0}^{n-1}\frac{t-l}{n-l} +
\sum_{n=1,3,5...}^\infty\frac{1-2\gamma}{(n+1)} \prod_{l=0}^{n-1}\frac{t-l}{n-l}
\end{align}
\textbf{Proof:}\ \ See Appendix~\ref{app_lem_ProbH4}.$\hfill\Box$
\end{lemma}
From Lemma~\ref{lem_ProbH3} and Lemma~\ref{lem_ProbH4}, we can  numerically compute the required number of measurements for any given $N$ and $K$. We will also provide an empirical study in Section~\ref{sec_exp}.

\section{Experiments and Comparisons }\label{sec_exp}

In this section, we provide a series of experimental studies to verify the proposed algorithm. In the  literature, the so-called 1-bit marginal regression~\cite{Article:Plan_IT13,Proc:Slawski_NIPS15} can be viewed as a one-scan algorithm and hence it is the  competitor we should compare our method with. As shown in the experiments, however, the proposed method needs orders of magnitude fewer measurements than 1-bit marginal regression.  Thus, to make the empirical study more interesting, we also compare the method with the well-known 1-bit Iterative Hard Thresholding (IHT)~\cite{Article:1Bit_IT13}.  The results can show that the proposed algorithm is still significantly more accurate. Furthermore,  our method is reasonably robust against random sign flipping, while IHT is known to be very sensitive to that kind of noise.

\subsection{A Practical Variant of Alg~\ref{alg_proposed}}

Although Alg.~\ref{alg_proposed} is convenient for theoretical analysis, the practical performance can be improved by using a simple variant based on ranking, although the theoretical analysis would be more difficult.

Basically, after we have computed $Q_i^+$ and $Q_i^-$ from (\ref{eqn_Q+}) and (\ref{eqn_Q-}), for $i =1$ to $N$, instead of using 0 as the threshold, we choose the top-$K$ coordinates ranked  by $\max\{Q_i^+, Q_i^-\}$. Among the selected coordinates, if $Q_i^+>Q_i^-$ (or $Q_i^->Q_i+$), then we estimate $sgn(x_i)$ to be positive (or negative). This procedure implicitly utilizes $\epsilon$ away from 0 and hence less conservative compared to vanilla  Alg.~\ref{alg_proposed}. In our experimental study, we always adopt this variant.

%top-$\beta K$ coordinates ranked by  $\max\{Q_i^+, Q_i^-\}$, for example, $\beta = 1.5$.

\subsection{Experiment Set-up}

In our experiments, we generate signals based on the two parameters $N$  and $K$. We choose $(N,K)\in\{(1000, 20), (1000,\ 50),  (10000,\ 20),  (10000,\ 50)\}$.  For each given $N$ and $K$, we first randomly select $K$ nonzero coordinates and  then assign the values  of the nonzero entries according to  i.i.d. samples from $N(0,5^2)$.   We then apply our proposed variant of Alg.~\ref{alg_proposed} to recover both the support and the signs of the signal.  The number of measurements is set according to
 \begin{align}\notag
 M = \zeta K \log N/\delta
 \end{align}
 where the confidence  $\delta$ is set to be 0.01. We vary the parameter  $\zeta$  from 2 to 15. Note that this choice of $M$ is typically a small number compared to $N$. Recall that, in our analysis, the required number of measurements using criterion (\ref{eqn_RecoveryQ0}) is proved to be $12.3K\log N/\delta$, although the actual measurements needed  will be  smaller.

\newpage

\subsection{Sign Recovery under Random Sign Flipping Noise}

\begin{figure}[h!]
\begin{center}
\mbox{
\includegraphics[width = 2.5in]{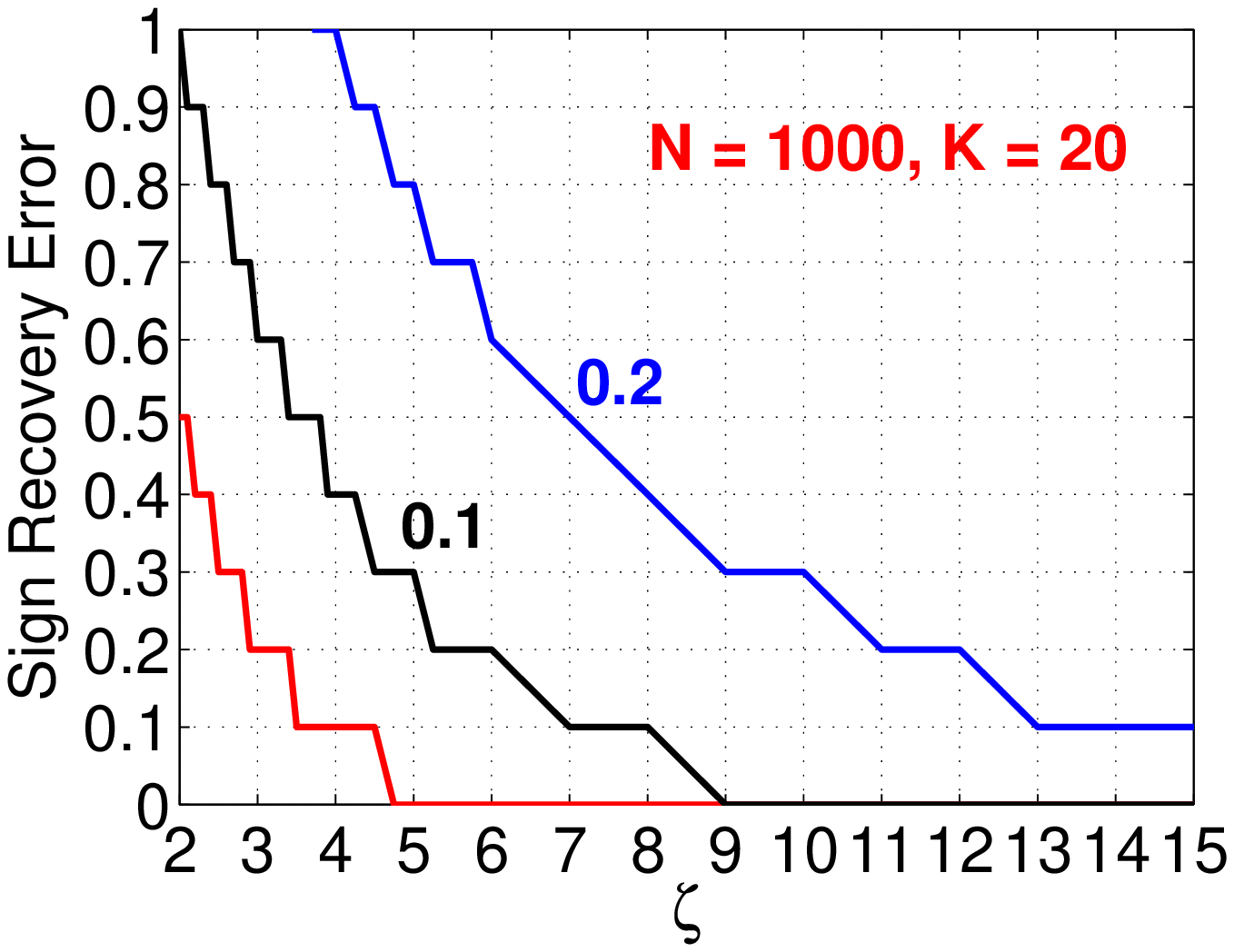}\hspace{0.3in}
\includegraphics[width = 2.5in]{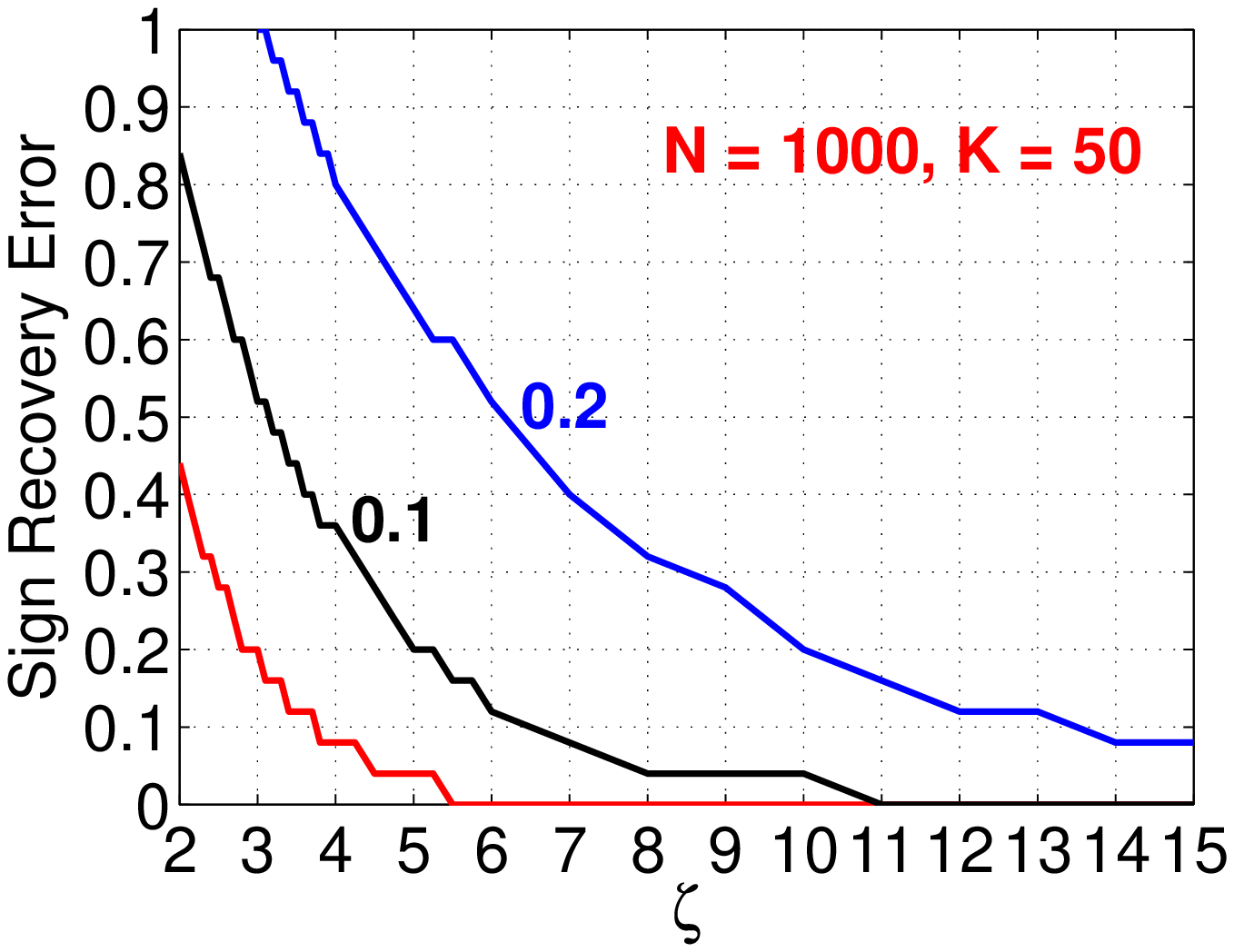}
}

\mbox{
\includegraphics[width = 2.5in]{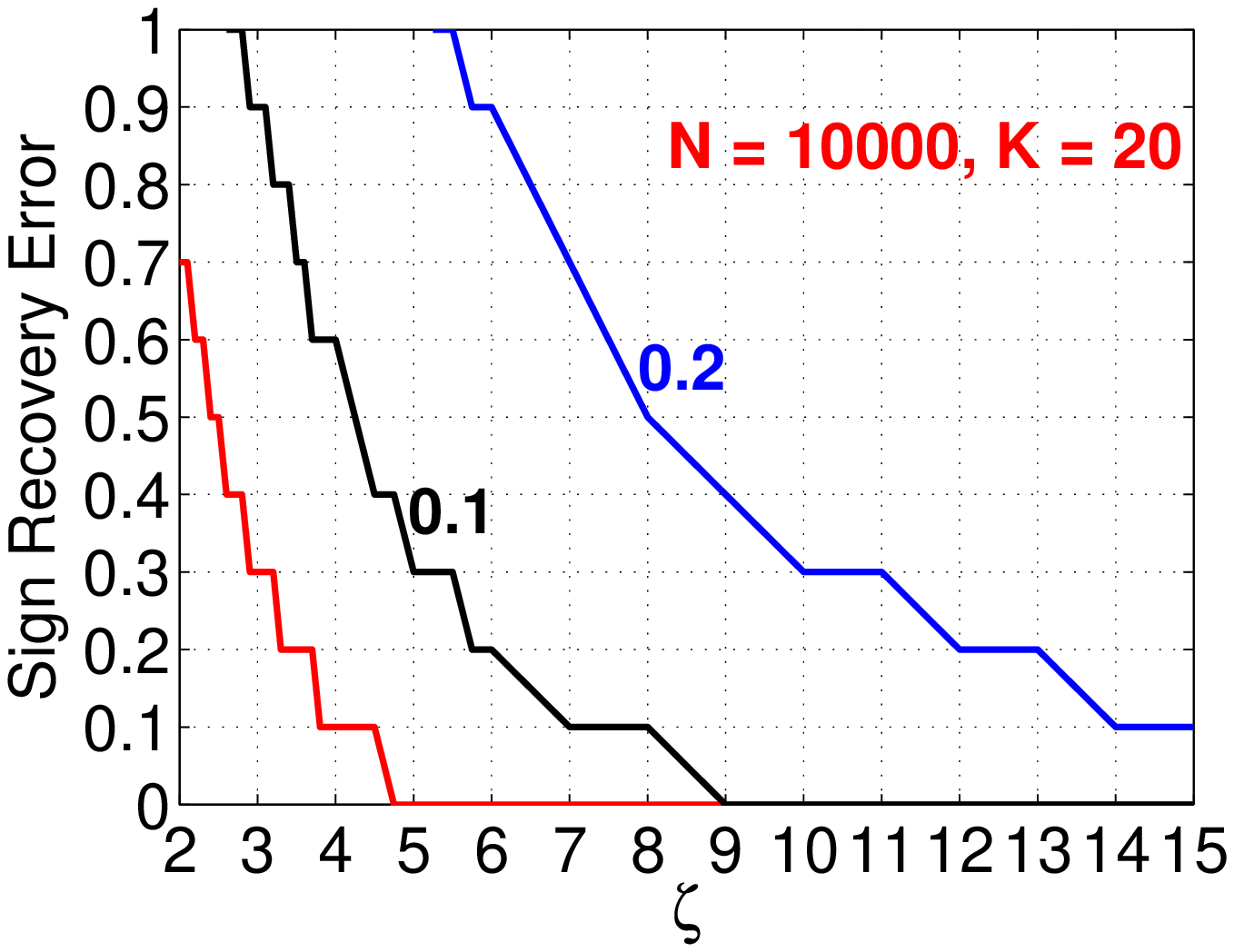}\hspace{0.3in}
\includegraphics[width = 2.5in]{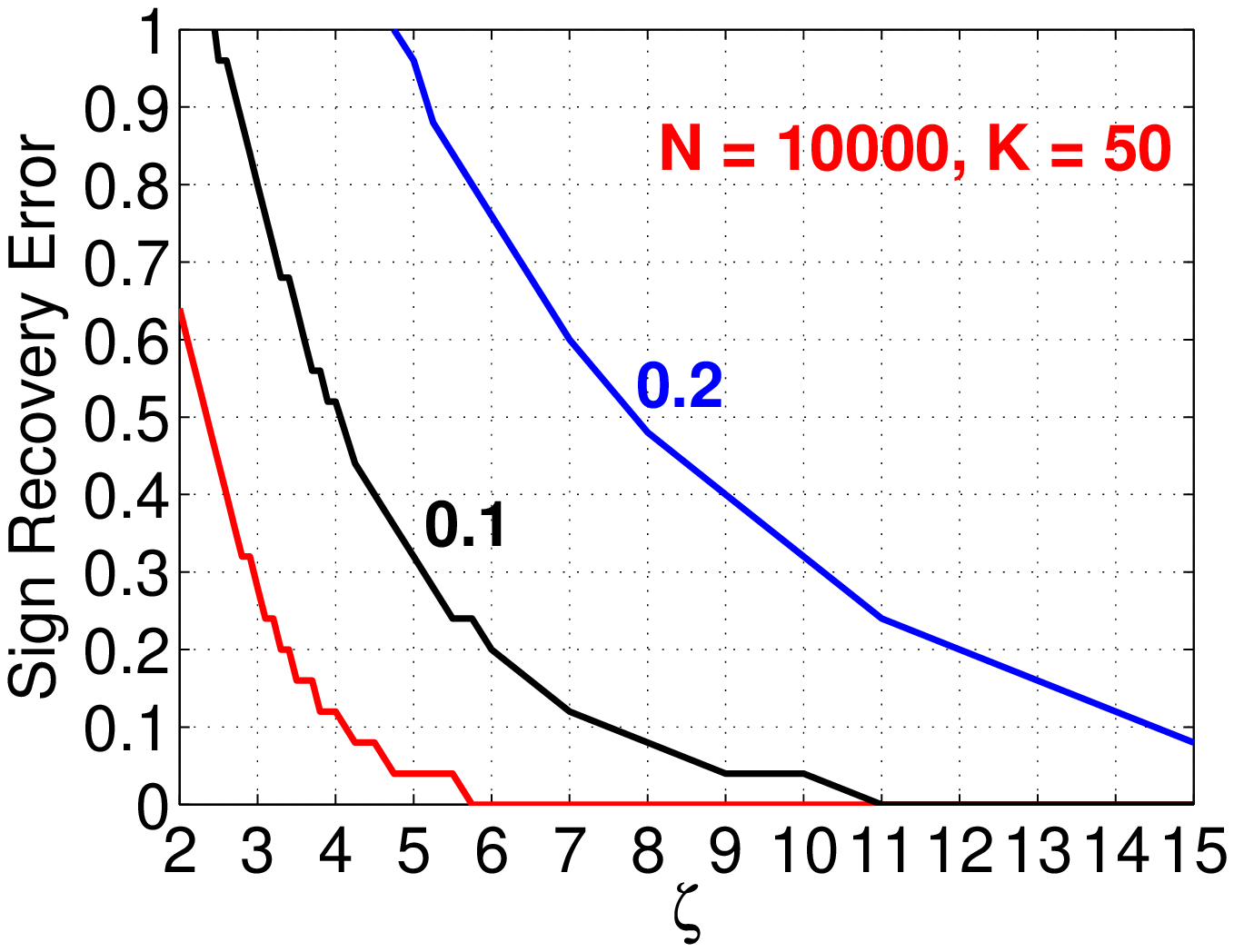}
}

\end{center}
\vspace{-0.2in}
\caption{\textbf{Sign recovery under random sign flipping noise}. The number of measurements is chosen according to $\zeta K\log N/\delta$, for $\zeta$ ranging from 2 to 15.    The recovery error is  $\sum_{i}|\hat{sgn(x_i)} - sgn(x_i)|/K$, where $i$ is from the top-$K$ reported coordinates  ranked by $\max\{Q_i^+, Q_i^-\}$. Note that using this definition,  the maximum possible sign recovery error is 2.  In each panel, the 3 curves correspond to 3 different random sign flipping probability $\gamma$, for $\gamma=0$, 0.1, and 0.2, respectively. The curve without label (red, if color is available) is for $\gamma =0$.  We repeat each simulation $1000$ times and report the medium.     } \label{fig_sign_flip_err}
\end{figure}

Figure~\ref{fig_sign_flip_err} reports the sign recovery errors $\sum_{i}|\hat{sgn(x_i)} - sgn(x_i)|/K$, where $i$ is from the top-$K$ reported coordinates. Note that using this definition,  the maximum sign recovery error can be as large as 2. In each panel, we report results for 3 different $\gamma$ values ($\gamma=0$, 0.1, and 0.2), where $\gamma$ is the random sign flipping probability. The curves without label (red, if color is available) correspond to $\gamma =0$ (i.e., no random sign flipping errors).\\

The results in Figure~\ref{fig_sign_flip_err} confirm that the proposed method works well as predicted by the theoretical analysis. Moreover, the method is fairly robust against random sign flipping noise.

\newpage
\subsection{Estimation of $K$ and the Impact on Recovery Performance}\label{sec_K_est}

In the theoretical analysis, we have assumed that $K$ is known, like many prior studies in compressed sensing. The problem becomes more interesting  when $K$ can not be assumed to be  known. In our framework, there are  two approaches to this problem. The first approach is to use a very small number of full measurements to estimate $K$.  Because the task of estimating $K$ is  much easier than the task of recovering the signal itself, it is reasonable to expect that the required number of measurements will be (very) small.
\\

\begin{figure}[h!]
\begin{center}
\mbox{
\includegraphics[width = 2.5in]{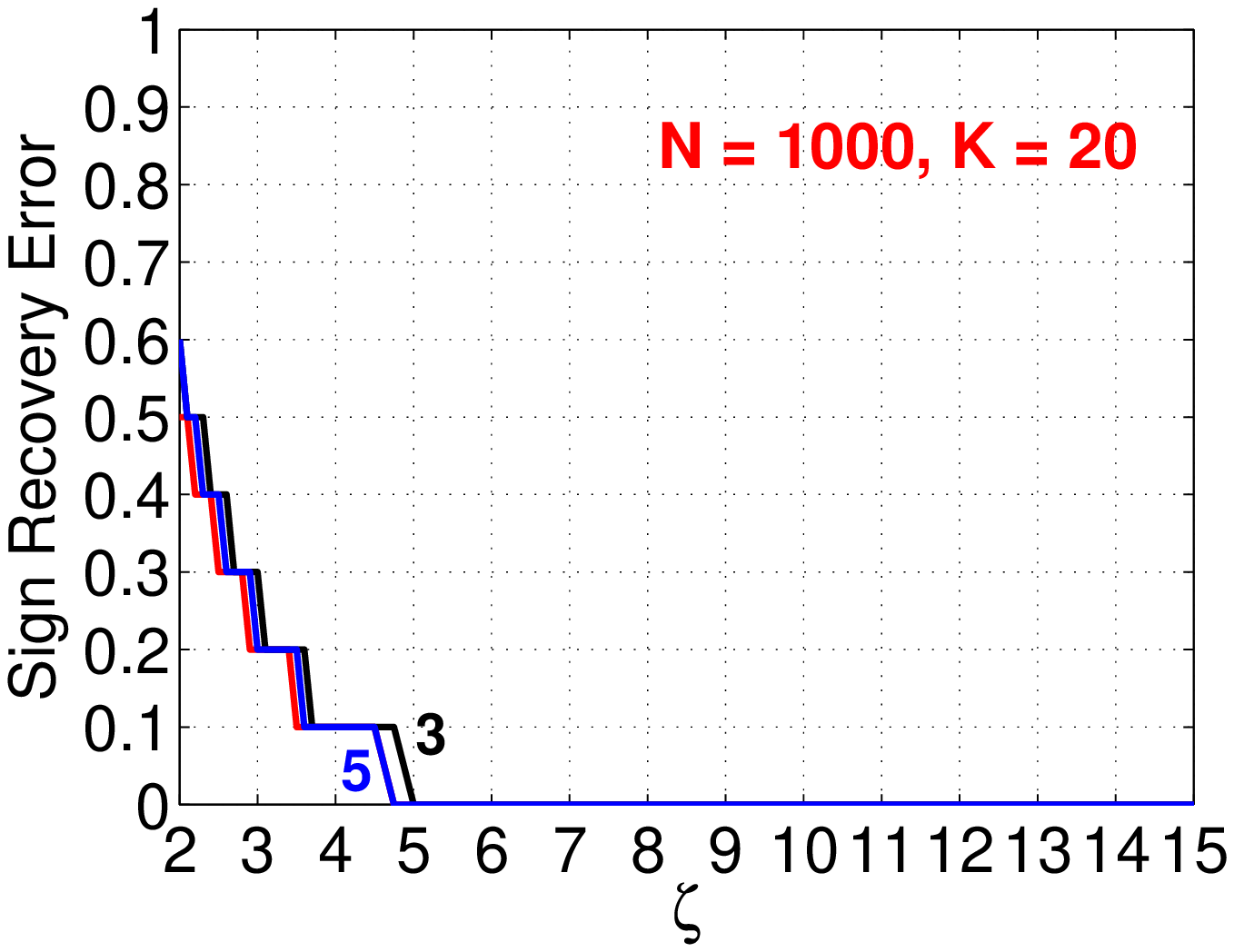}\hspace{0.3in}
\includegraphics[width = 2.5in]{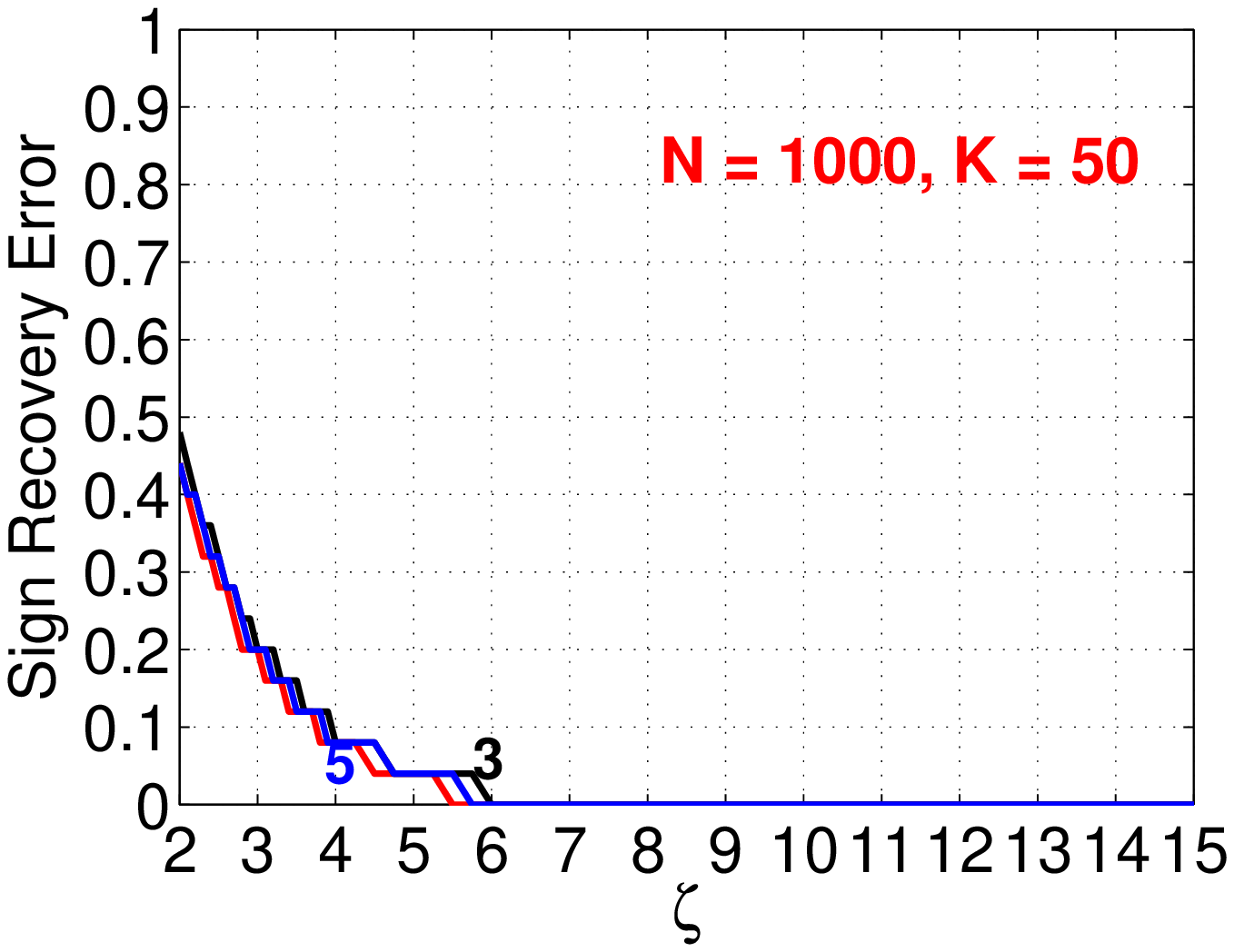}
}

\mbox{
\includegraphics[width = 2.5in]{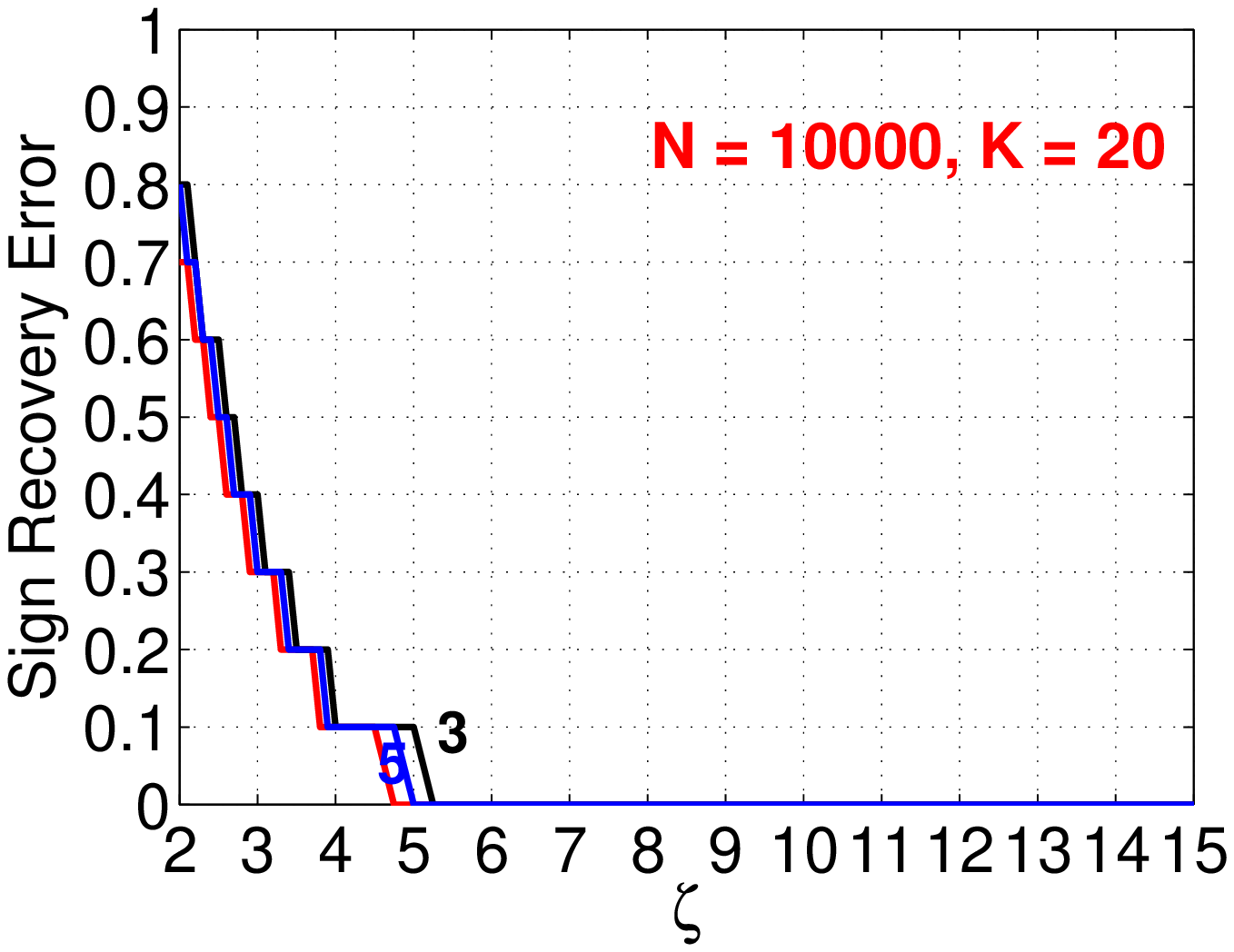}\hspace{0.3in}
\includegraphics[width = 2.5in]{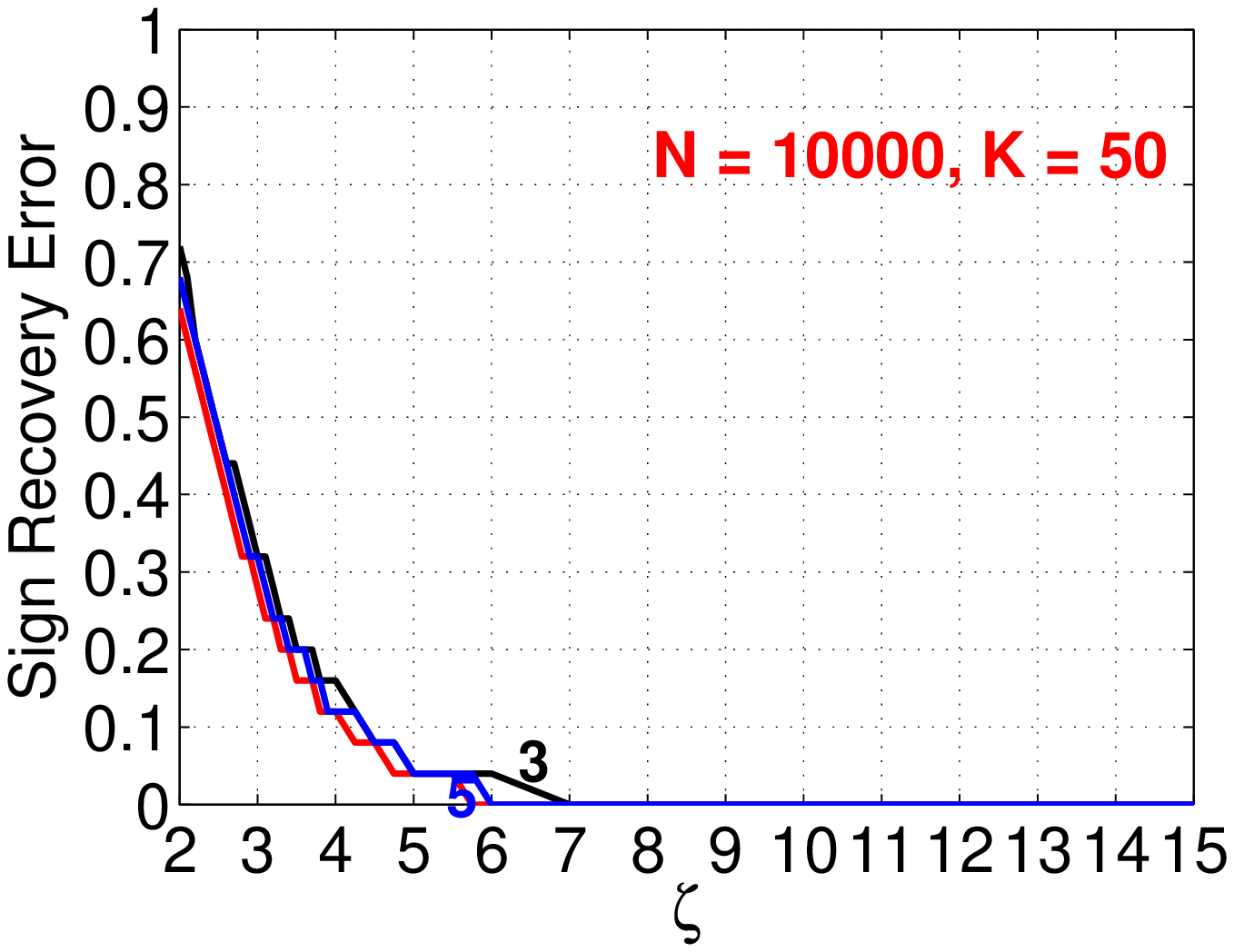}
}

\end{center}
\vspace{-0.2in}
\caption{\textbf{Sign recovery with estimated $K$} by the harmonic mean estimator~\cite{Proc:Li_SODA08}. In each panel, the unlabeled curve (red if color is available)  corresponds to the use of exact values of $K$. With merely 5 samples (curves labeled ``5'') for estimating $K$, the recovery results are already  close to results using  exact $K$ values.    } \label{fig_sign_err_Khat}
\end{figure}

 Here we use the harmonic mean estimator~\cite{Proc:Li_SODA08}:
\begin{align}\label{eqn_Khat}
&\hat{K} = \frac{-\frac{2}{\pi}\Gamma(-\alpha)\sin\frac{\pi}{2}\alpha}{\sum_{j=1}^M\frac{1}{|y_j|^\alpha}}\left(M-\left(\frac{-\pi\Gamma(-2\alpha)\sin(\pi\alpha)}{\left[\Gamma(-\alpha)\sin\frac{\pi}{2}\alpha\right]^2}-1\right)\right)
\end{align}
For small $\alpha$, $\hat{K}$ is essentially $M/\sum_{j=1}^M 1/|y_j|^\alpha$ with the variance essentially being $\frac{K^2}{M}$.  Figure~\ref{fig_sign_err_Khat} provides a  set of experiments to confirm that only using a very small number (such as 5) of measurements to estimate $K$ leads to very accurate results, compared to using the exact values of $K$.\\

Another line of approach is to develop  bit-estimators of $K$, which is an interesting and separate research problem, as reported in~\cite{Report:SymStableCode}

\newpage

\subsection{Support Recovery}

\begin{figure}[h!]
\begin{center}
\mbox{
\includegraphics[width = 2.5in]{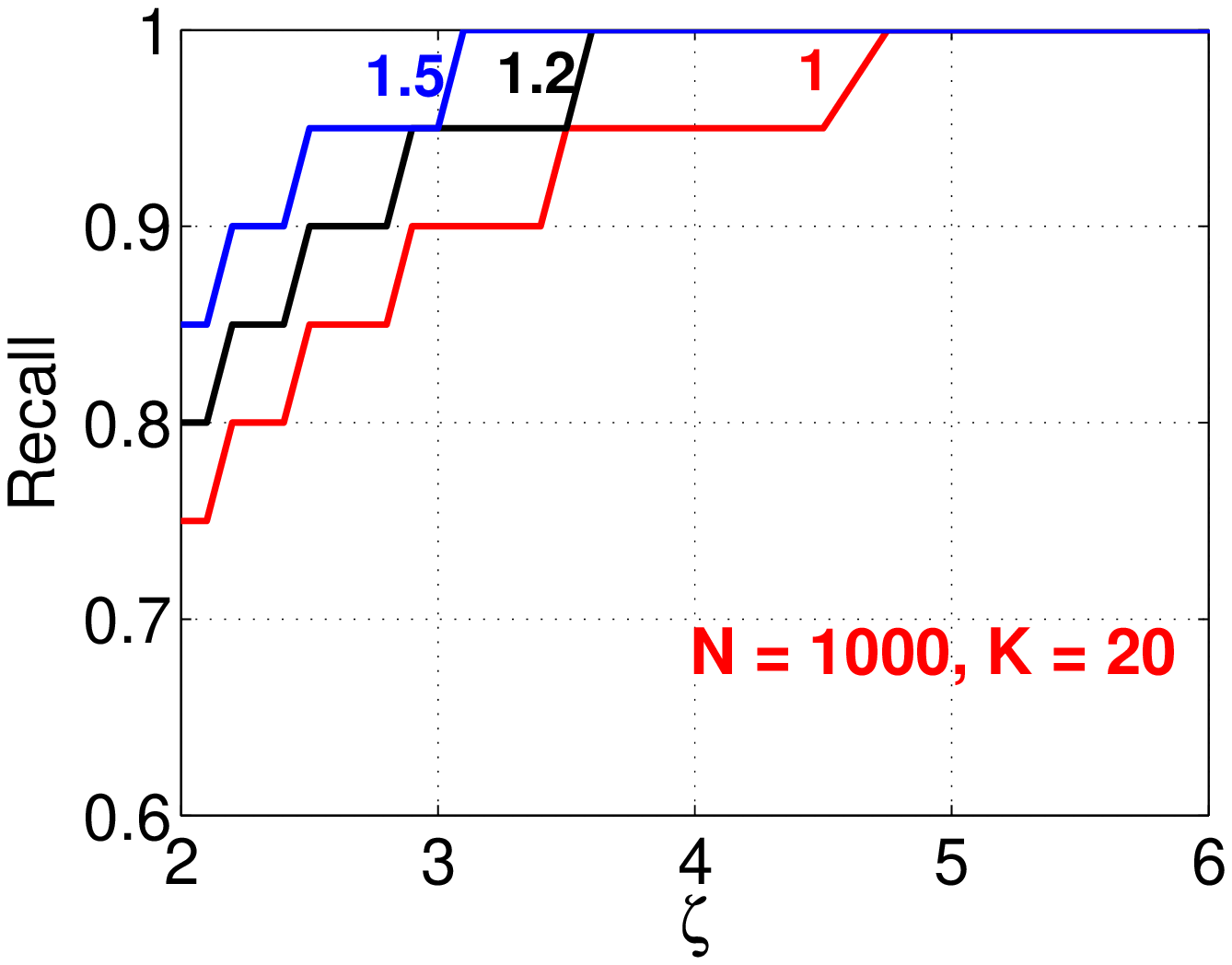}\hspace{0.3in}
\includegraphics[width = 2.5in]{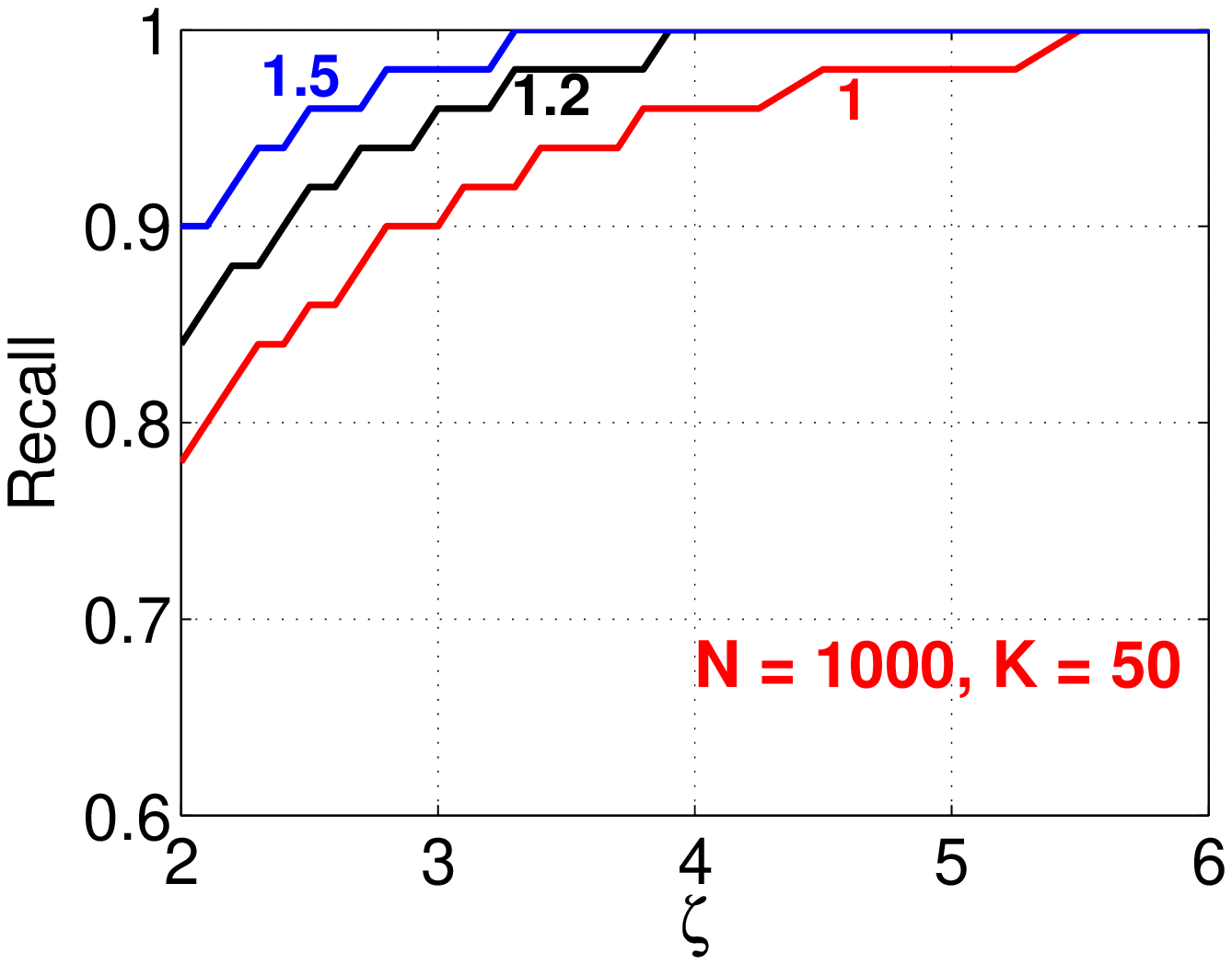}
}

\mbox{
\includegraphics[width = 2.5in]{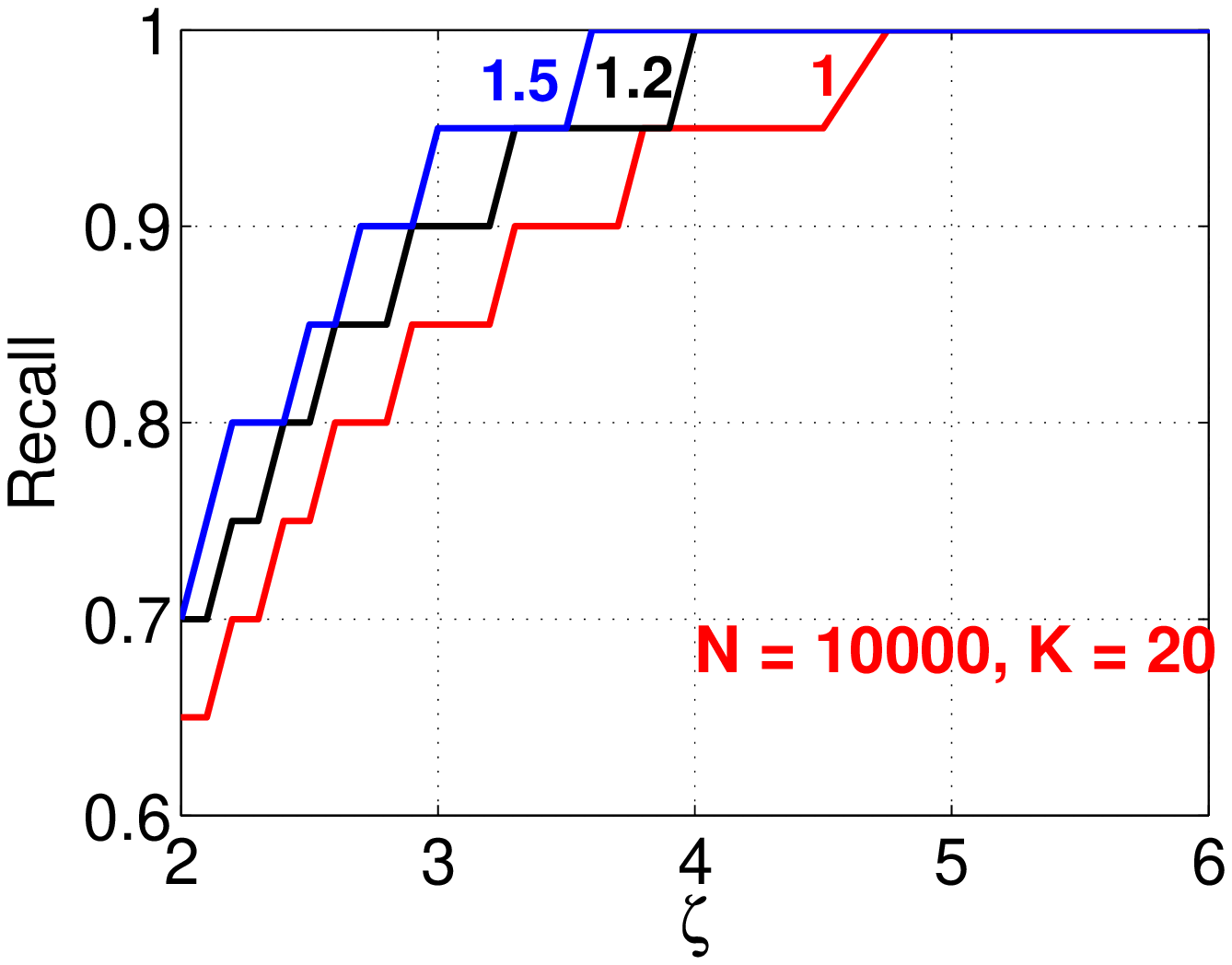}\hspace{0.3in}
\includegraphics[width = 2.5in]{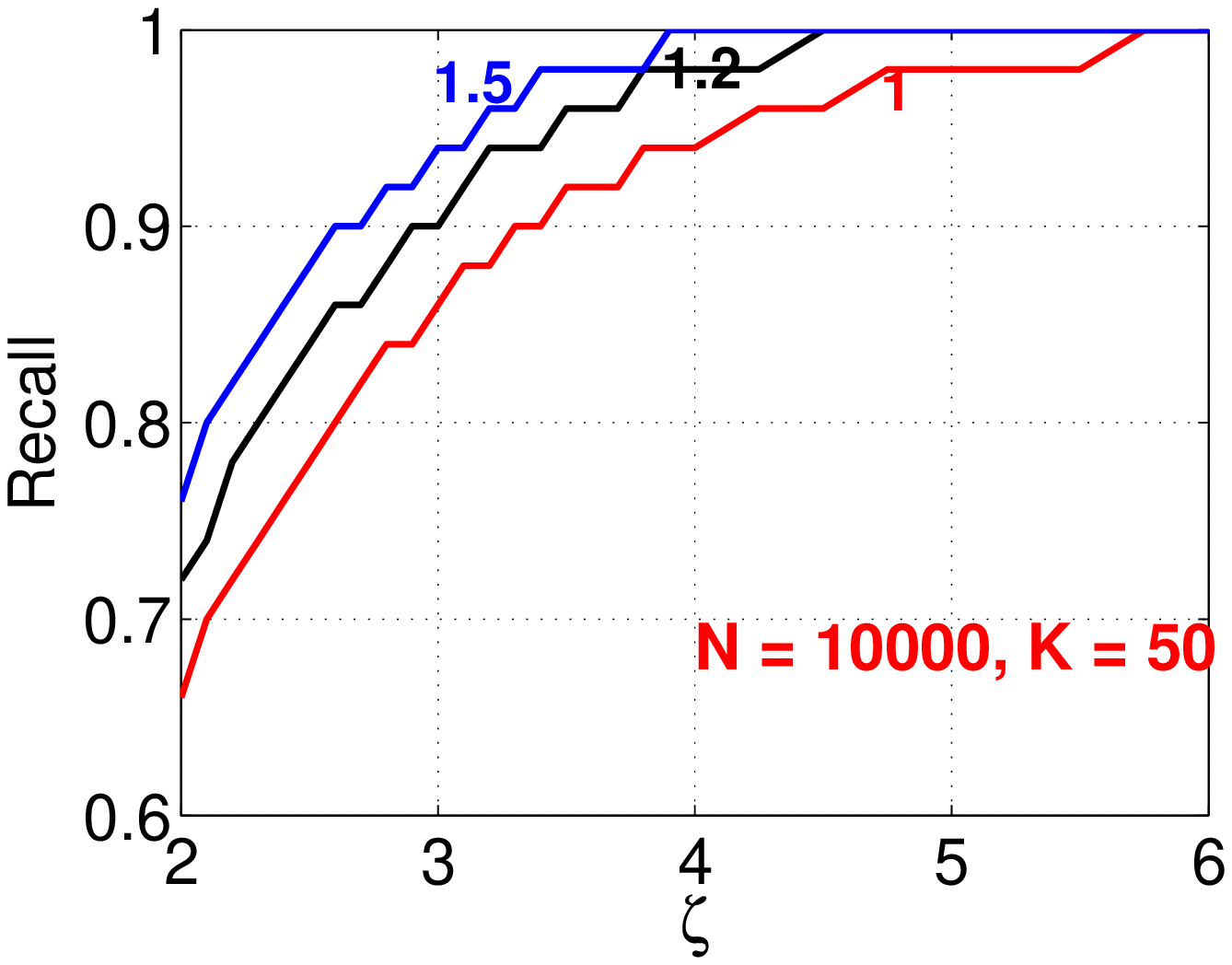}
}

\end{center}
\vspace{-0.2in}
\caption{\textbf{Support recovery}. We report top-$\beta K$ coordinates ranked by $\max\{Q_i^+, Q_i^-\}$, for $\beta\in\{1, 1.2, 1.5, 2\}$. We report the \textbf{recall} values, i.e., $\#\{retrieved\ true\ nonzeros\}/K$. As expected, using larger $\beta$ will reduce the required number of measurements, which is set to be $\zeta K\log N/\delta$ (where $\delta=0.01$). } \label{fig_recall}
\end{figure}

We can generalize the practical variant of Alg.~\ref{alg_proposed}. That is, after we rank the coordinates according to $\max\{Q_i^+, Q_i^-\}$, we can choose top-$\beta K$ coordinates for $\beta\geq 1$. We have used $\beta=1$ in previous experiments.  Figure~\ref{fig_recall}  reports the recall values for support recovery:
\begin{align}\notag
recall = \#\{retrieved\ true\ nonzeros\}/K
\end{align}
for $\beta=1$, 1.2, and 1.5. Note that in this case we just need to present the recalls,  because $precision = \#\{retrieved\ true\ nonzeros\}/(\beta K)$. \\

As expected, using larger $\beta$ values can reduce the required number of measurements. This experiment could be interesting for practitioners who care about this trade-off.

\newpage

\subsection{Comparisons with 1-bit Marginal Regression}

It is helpful to provide a comparison study with other 1-bit algorithms in the literature. Unfortunately, most of those available 1-bit algorithms are not one-scan methods. One exception is the 1-bit marginal regression~\cite{Article:Plan_IT13,Proc:Slawski_NIPS15}, which can be viewed as a one-scan algorithm.  Thus, it is the target competitor we should compare our method with. \\

\begin{figure}[h!]
\begin{center}
\mbox{
\includegraphics[width = 2.5in]{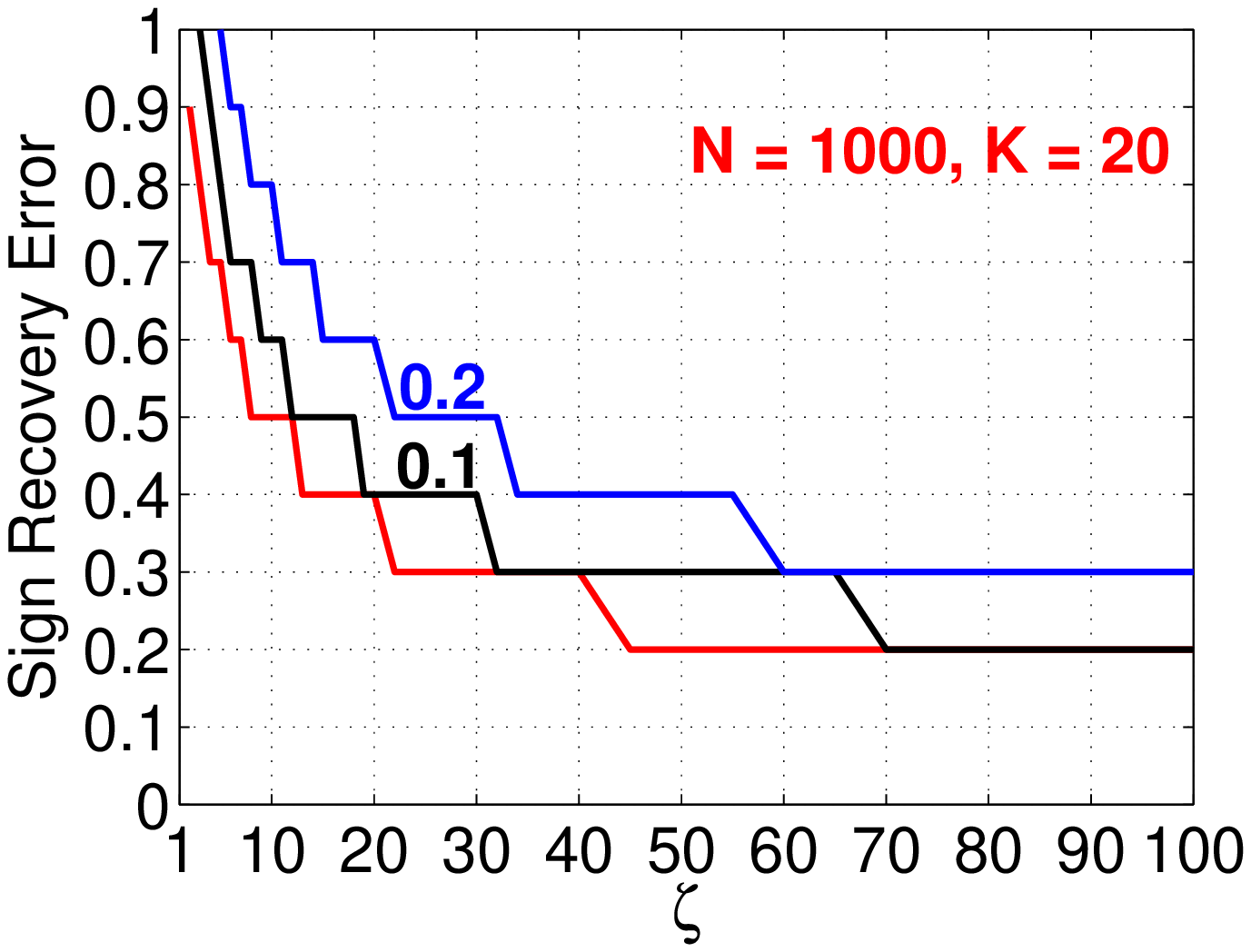}\hspace{0.3in}
\includegraphics[width = 2.5in]{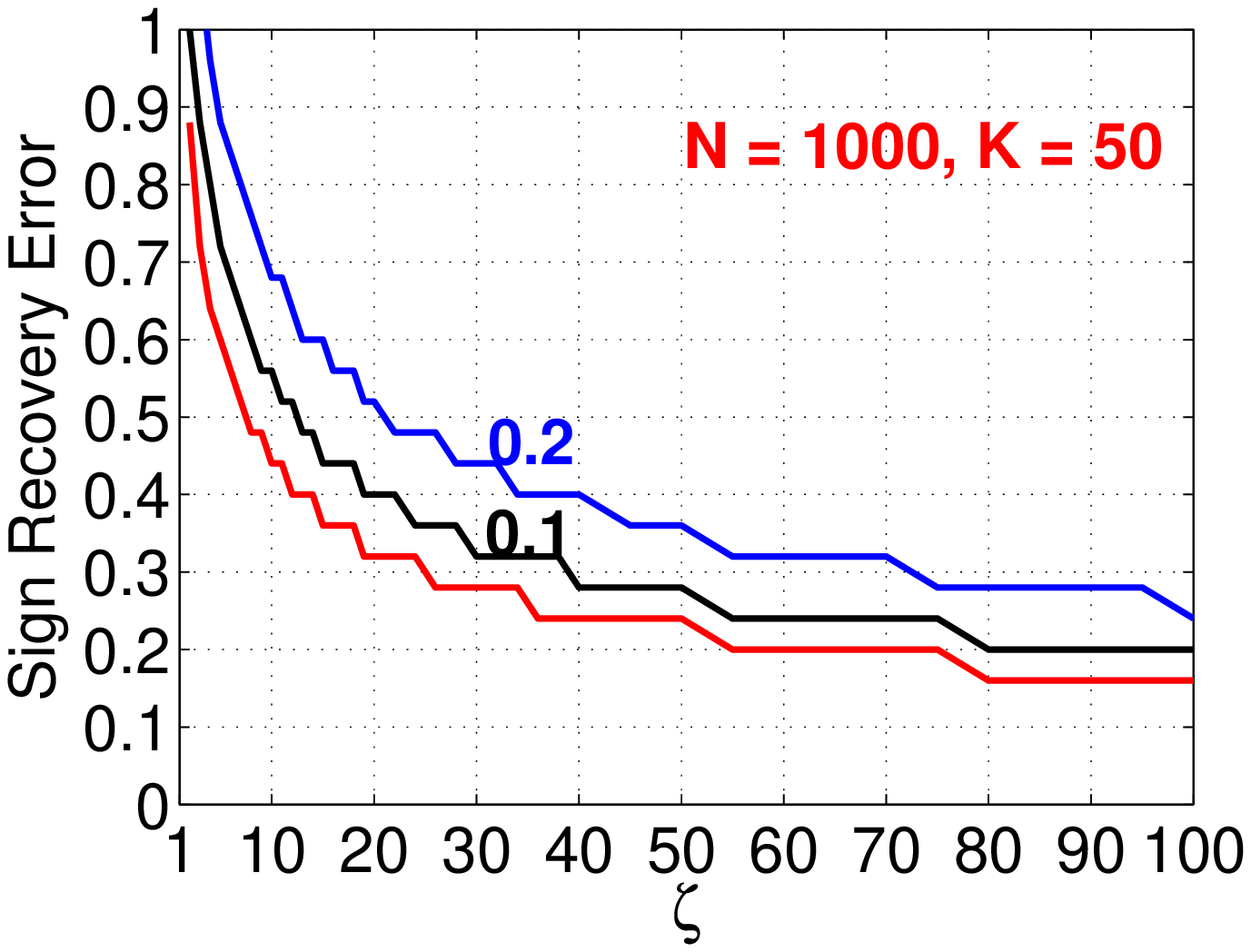}
}
\mbox{
\includegraphics[width = 2.5in]{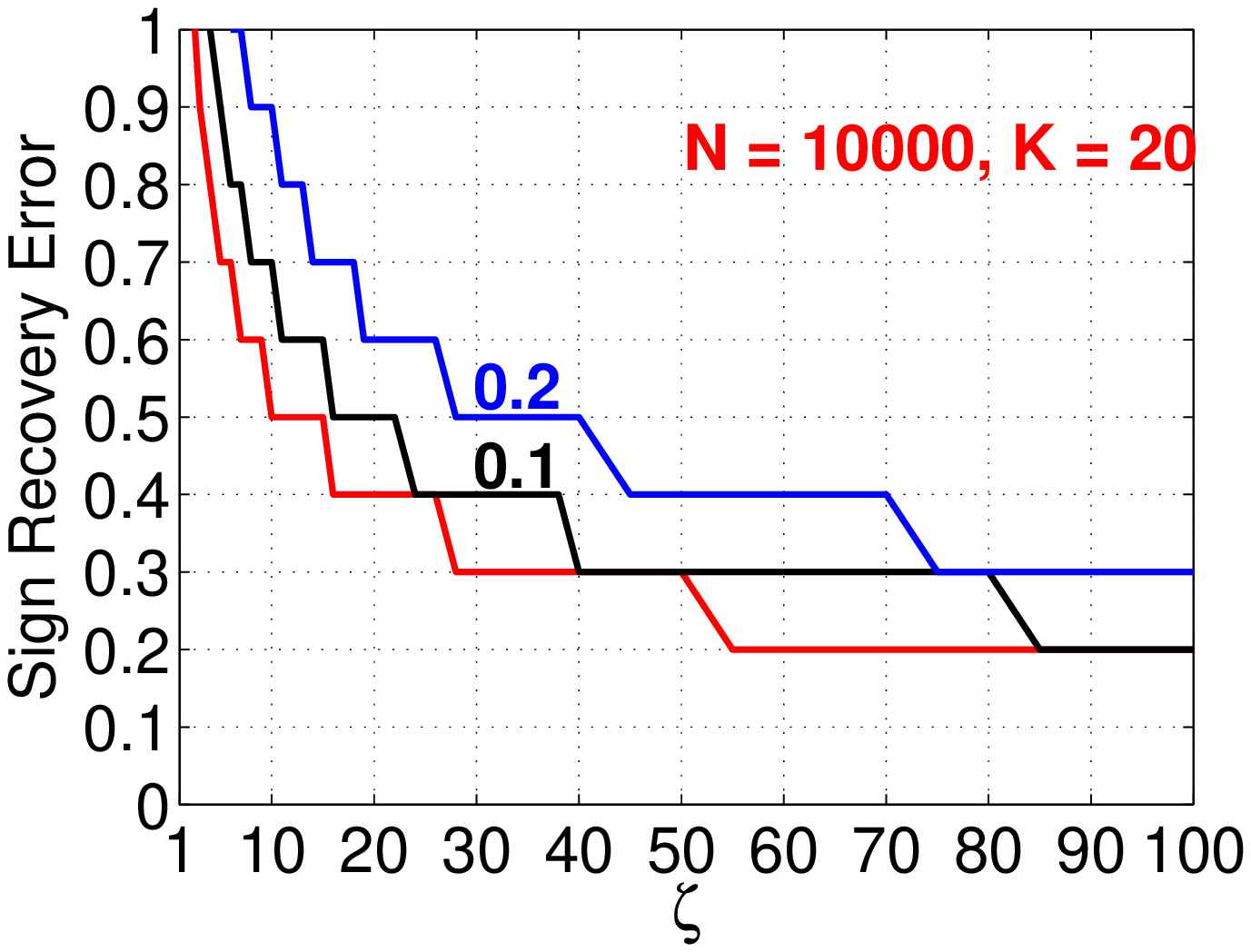}\hspace{0.3in}
\includegraphics[width = 2.5in]{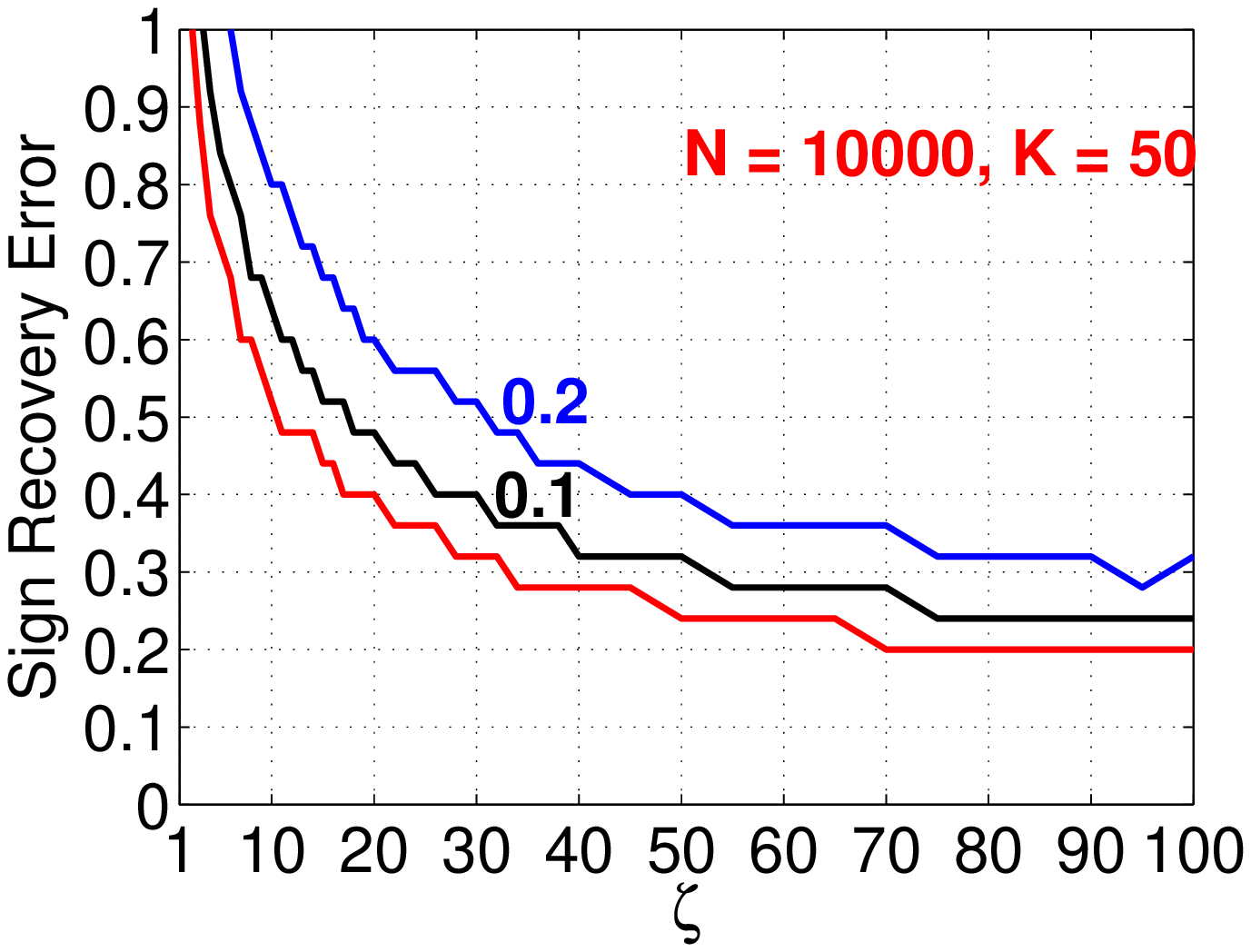}
}

\end{center}
\vspace{-0.2in}
\caption{\textbf{Sign recovery with 1-bit marginal regression}. The errors are still very larger even with $\zeta=100$, i.e., $M=100K\log N/\delta$. Note that in each panel, the three curves correspond to three different random sign flipping probabilities: $\gamma = 0$, 0.1, and 0.2, respectively.  }\label{fig_MR}
\end{figure}

Figure~\ref{fig_MR} reports the sign recovery accuracy of 1-bit marginal regression in our experimental setting. That is, we also choose $M = \zeta K\log N/\delta$, although for this approach, we must enlarge $\zeta$ dramatically, compared to our proposed method. We can see that even with $\zeta=100$, the errors of 1-bit marginal regression are still  large.

\newpage

\subsection{Comparisons with 1-bit Iterated Hard Thresholding (IHT)}

\begin{figure}[h!]
\begin{center}
\mbox{
\includegraphics[width = 2.5in]{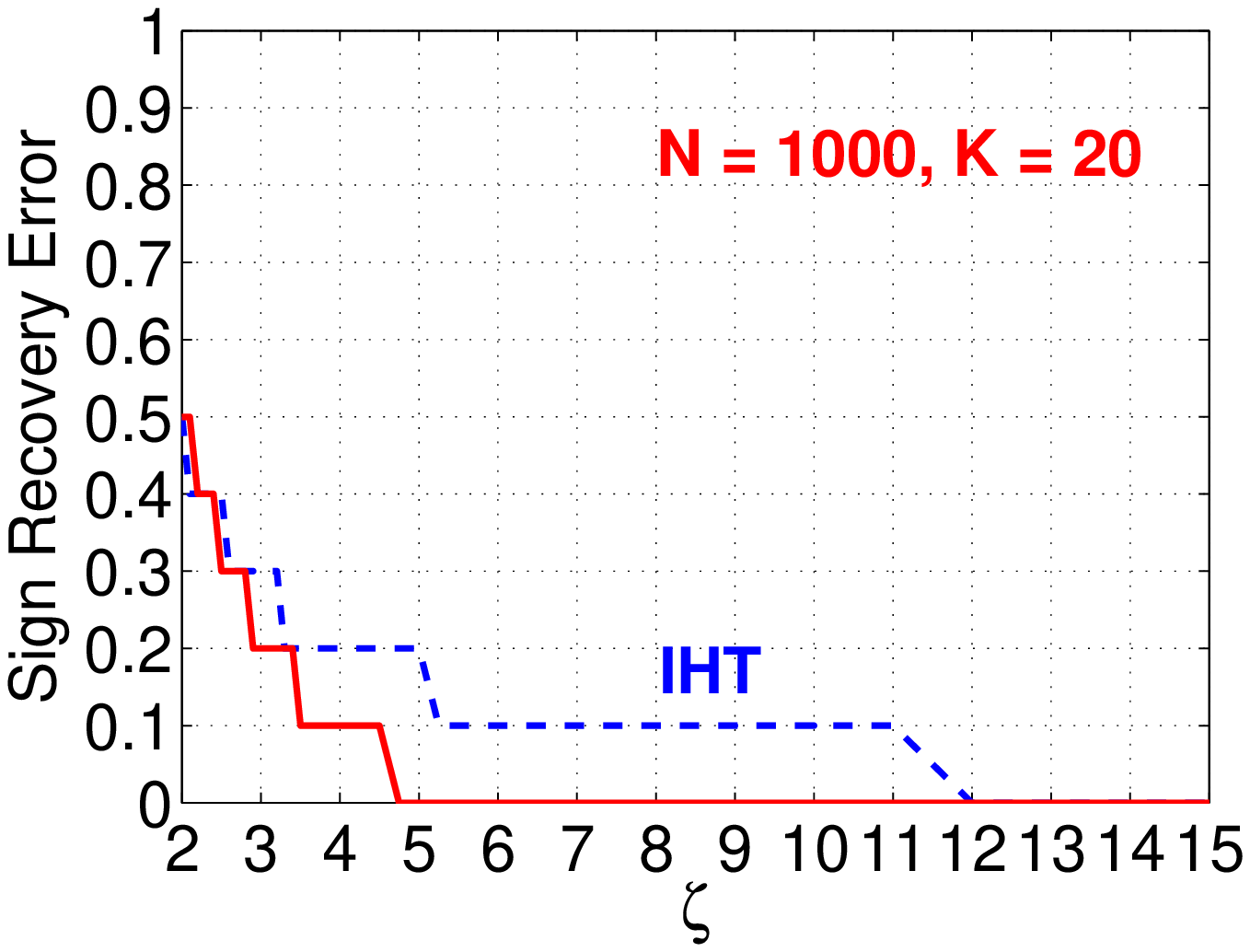}\hspace{0.3in}
\includegraphics[width = 2.5in]{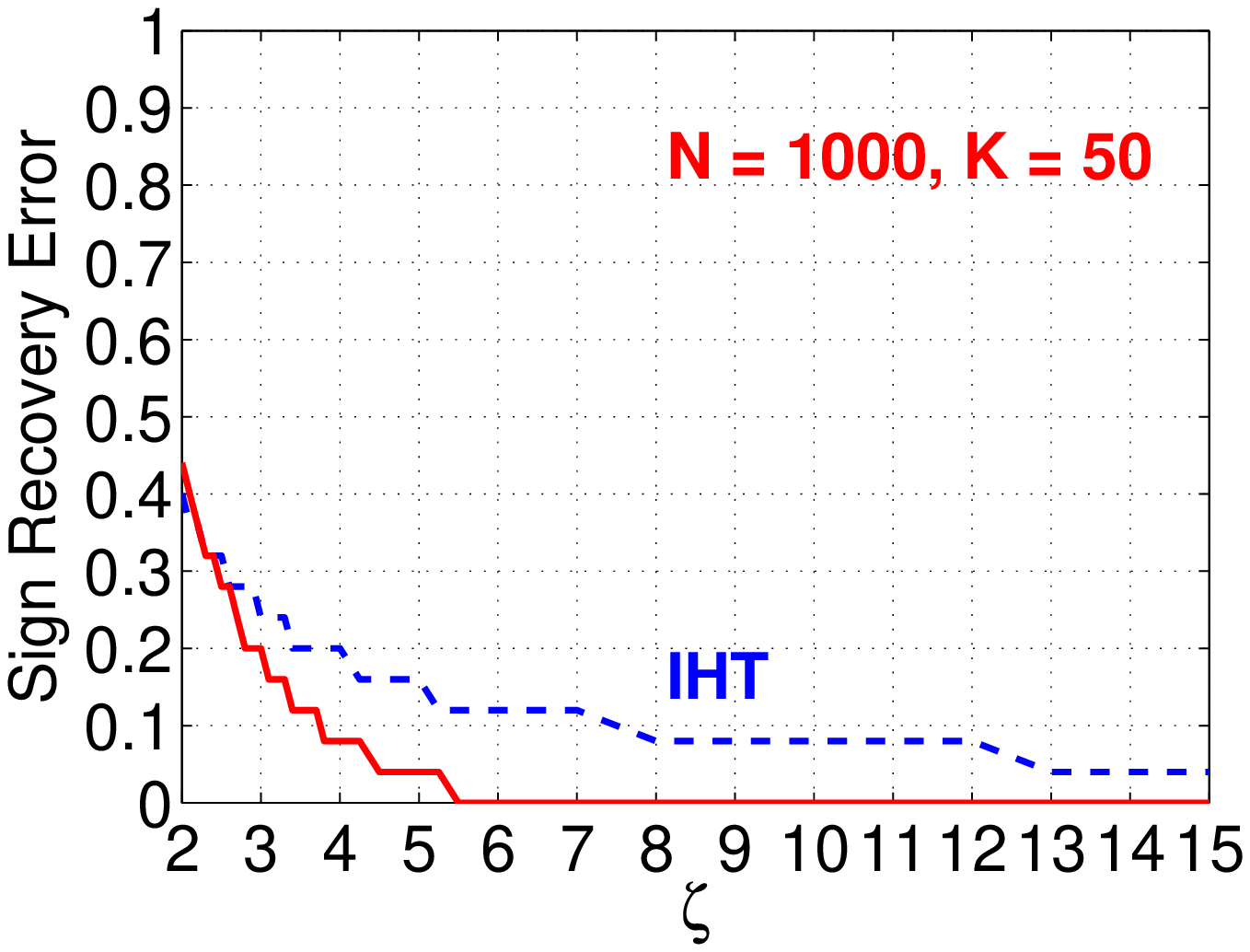}
}
\mbox{
\includegraphics[width = 2.5in]{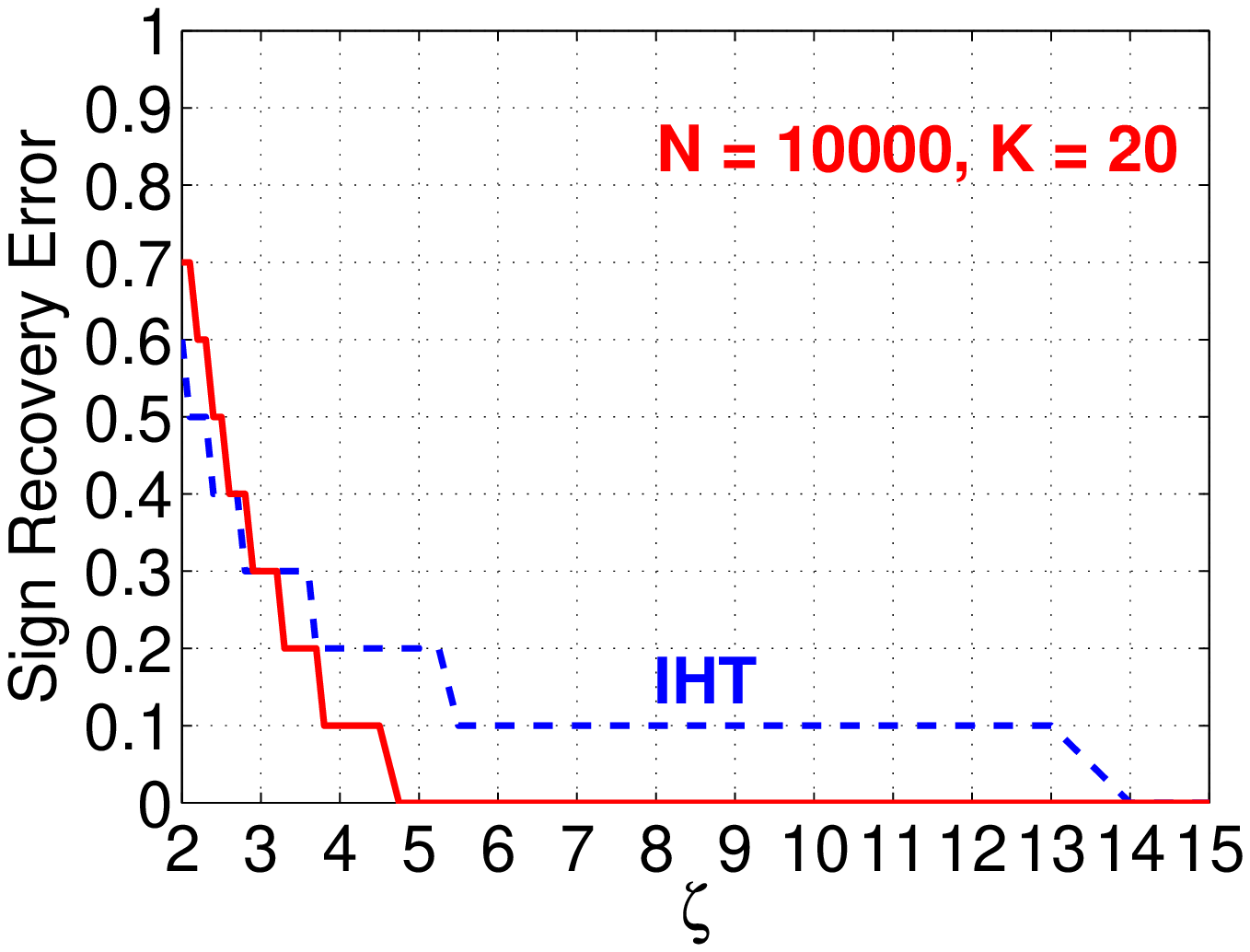}\hspace{0.3in}
\includegraphics[width = 2.5in]{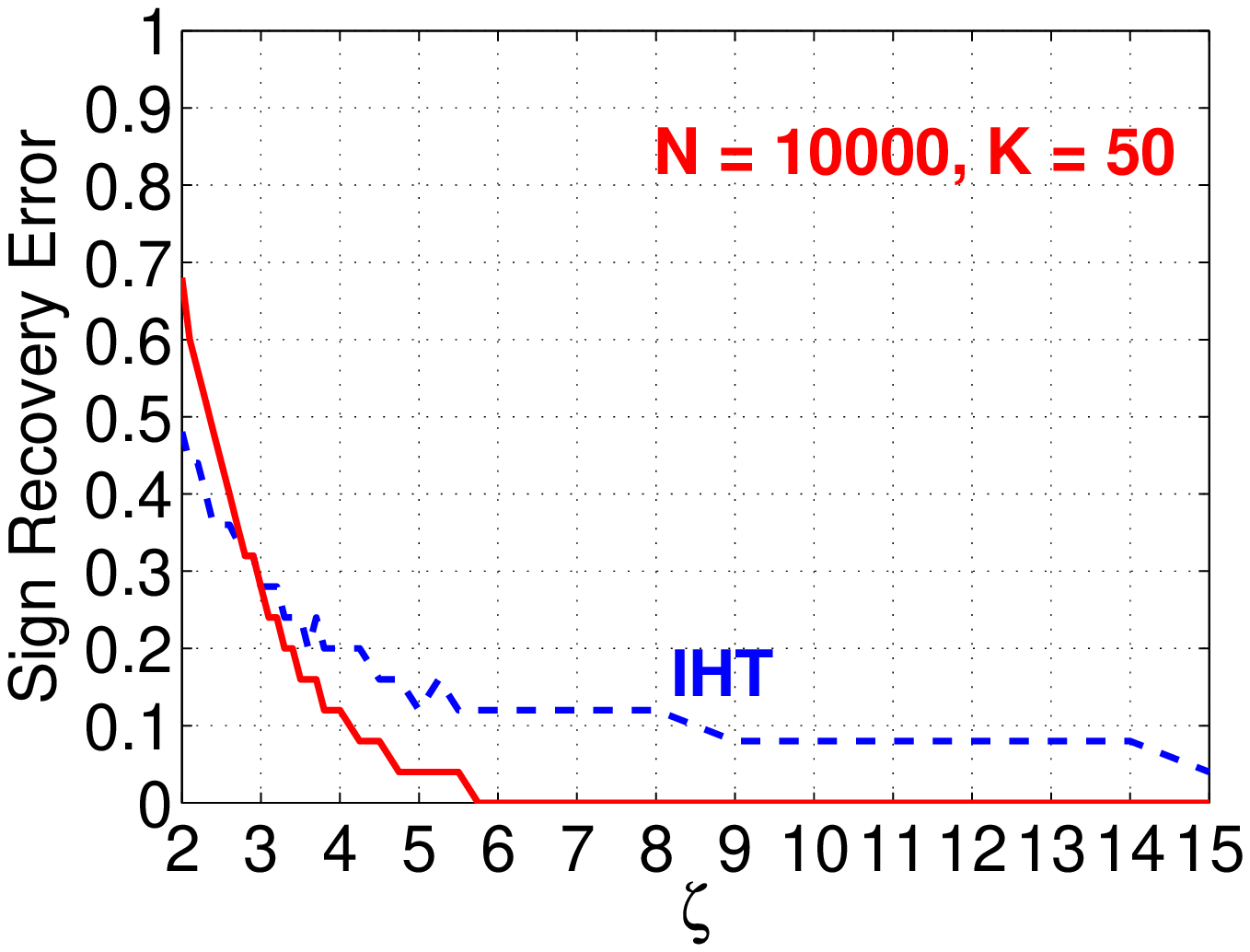}
}

\end{center}
\vspace{-0.2in}
\caption{\textbf{Sign recovery with 1-bit iterative hard thresholding (IHT)}. The results of 1-bit IHT are presented as dashed (blue, if color is available) curves. For comparison, we also plot the results of the proposed method (solid and red if color is available).    } \label{fig_IHT}\vspace{-0.2in}
\end{figure}

We conclude this section by providing a comparison with the well-known 1-bit iterative hard thresholding (IHT)~\cite{Article:1Bit_IT13}.  Even though 1-bit IHT is not a one-scan algorithm, we compare it with our method for completeness. As shown in Figure~\ref{fig_IHT}, the proposed algorithm is still significantly more accurate for sign recovery. 

Note that Figure~\ref{fig_IHT} does not include results of 1-bit IHT with random sign flipping noise. As previously shown, the proposed method is reasonably robust against this type of noise. However, we observe that 1-bit IHT is so sensitive to random sign flipping that the results are not presentable~\footnote{After consulting the author of~\cite{Article:1Bit_IT13}, we decided not to present the random sign flipping experiment for 1-bit IHT.}.

\section{Conclusion}

1-bit compressed sensing (CS) is an important topic because the measurements are typically quantized (by hardware) and using only the sign information may potentially lead to cost reduction in collection, transmission, storage, and retrieval. Current methods for 1-bit CS are less satisfactory because they require a very large number of measurements and the decoding is typically not one-scan. Inspired by recent method of compressed sensing with very heavy-tailed design, we develop an algorithm for one-scan 1-bit CS, which is provably accurate and fast, as validated by experiments.\\

\noindent For sign recovery, our proposed one-scan 1-bit algorithm requires orders of magnitude fewer measurements compared to 1-bit marginal regression.  Our method is still significantly more accurate than 1-bit Iterative Hard Thresholding (IHT), which is not one-scan. Moreover, unlike 1-bit IHT,  the proposed algorithm is reasonably robust again random sign flipping noise.

\vspace{0.2in}
%\newpage \clearpage

\appendix

\noindent\textbf{Appendix}

\section{Proof of Lemma~\ref{lem_ProbH1}}\label{app_lem_ProbH1}
Recall
\begin{align}\notag
Q_i^{+} = \sum_{j=1}^M\log \left(1+{sgn(y_j)}{sgn(u_{ij})}e^{-\left({K}-1\right)w_{ij}}\right)
= \sum_{j=1}^M\log \left(1+sgn\left({y_j}/{s_{ij}}\right)e^{-\left({K}-1\right)w_{ij}}\right)
\end{align}
where  $\frac{y_j}{s_{ij}} = x_i + \frac{\sum_{t\neq i} x_i s_{tj}}{s_{ij}} = x_i +\theta_i \frac{S_j}{s_{ij}}$. Here,  $S_j\sim S(\alpha,1)$ is independent of $s_{ij}$, and for convenience we define $\theta = \left(\sum_{i=1}^N |x_i|^\alpha\right)^{1/\alpha}$ and  $\theta_i = \left(\theta^\alpha- |x_i|^\alpha\right)^{1/\alpha}$. In particular, if $x_i=0$, then $\theta_i = \theta$ and $sgn\left({y_j}/{s_{ij}}\right) =  sgn(S_j/s_{ij})$. As $S_j$ and $s_{ij}$ are symmetric and independent,  we can replace $sgn(S_j/s_{ij})$ by $sgn(s_{ij})=sgn(u_{ij})$. To see this
\begin{align}\notag
&\mathbf{Pr}\left(sgn(S_j/s_{ij}) = 1\right) = \mathbf{Pr}\left(sgn(s_{ij}/S_{j})=1\right)\\\notag
=& \mathbf{Pr}\left(sgn(s_{ij})=1\right)\mathbf{Pr}\left(S_j>0\right)+\mathbf{Pr}\left(sgn(s_{ij})=-1\right)\mathbf{Pr}\left(S_j<0\right)\\\notag
 =& \frac{1}{2}\frac{1}{2}+\frac{1}{2}\frac{1}{2}=\frac{1}{2} =\mathbf{Pr}\left(sgn(s_{ij})=1\right)
\end{align}
Thus, we have
\begin{align}\notag
&\mathbf{Pr}\left(Q_i^+>\epsilon M/K, x_i=0\right)\\\notag
=&\mathbf{Pr}\left(\sum_{j=1}^M\log \left(1+sgn(y_j/s_{ij})\exp\left(-(K-1)w_{ij}\right)\right)>\epsilon M/K,x_i=0\right)\\\notag
=&\mathbf{Pr}\left(\sum_{j=1}^M\log \left(1+sgn(S_j/s_{ij})\exp\left(-(K-1)w_{ij}\right)\right)>\epsilon M/K\right)\\\notag
=&\mathbf{Pr}\left(\sum_{j=1}^M\log \left(1+sgn(u_{ij})\exp\left(-(K-1)w_{ij}\right)\right)>\epsilon M/K\right)\\\notag
=&\mathbf{Pr}\left(\prod_{j=1}^M \left(1+sgn(u_{ij})\exp\left(-(K-1)w_{ij}\right)\right)>e^{\epsilon M/K}\right)\\\notag
\leq&e^{-\epsilon M/K t} E^M\left(1+sgn(u_{ij})\exp\left(-(K-1)w_{ij}\right)\right)^t,\hspace{0.5in} (t\geq 0, \text{Markov's Inequality}) \\\notag
=&e^{-\epsilon M/K t}\left(\frac{1}{2}E\left\{\left(1+e^{-(K-1)w_{ij}}\right)^t +  \left(1-e^{-(K-1)w_{ij}}\right)^t\right\}\right)^M\\\notag
=&e^{-\epsilon  M/K t}\left(\frac{1}{2}\int_0^\infty\left\{\left(1+e^{-(K-1)w}\right)^t +  \left(1-e^{-(K-1)w}\right)^t \right\}e^{-w}dw\right)^M
\end{align}
Then we need to choose the $t$ to minimize the upper bound. Let $b=K-1$, then
\begin{align}\notag
&\int_0^\infty \left(1+e^{-bw}\right)^t e^{-w}dw =  \int_0^1(1+u^b)^t du
\\\notag
=&\int_0^1 1+u^bt + u^{2b}t(t-1)/2! + u^{3b}t(t-1)(t-2)/3! + u^{4b}t(t-1)(t-2)(t-3)/4! + .... d u\\\notag
=&1 + \frac{t}{b+1} + \frac{t(t-1)}{(2b+1)2!} +  \frac{t(t-1)(t-2)}{(3b+1)3!}+...
\end{align}

\begin{align}\notag
&\int_0^\infty \left(1-e^{-bw}\right)^t e^{-w}dw =  \int_0^1(1-u^b)^t du\\\notag
=&\int_0^1 1-u^bt + u^{2b}t(t-1)/2! -u^{3b}t(t-1)(t-2)/3! + u^{4b}t(t-1)(t-2)(t-3)/4! + .... d u\\\notag
=&1 -\frac{t}{b+1} + \frac{t(t-1)}{(2b+1)2!} - \frac{t(t-1)(t-2)}{(3b+1)3!}+...
\end{align}
\begin{align}\notag
&\int_0^\infty \left(1-e^{-(K-1)w}\right)^t e^{-w} + \left(1+e^{-(K-1)w}\right)^t e^{-w}dw
=2  + 2\frac{t(t-1)}{(2K-1)2!} +  2\frac{t(t-1)(t-2)(t-3)}{(4K-3)4!}+...
\end{align}
Therefore, for any $t\geq 0$, we have
\begin{align}\notag
&\mathbf{Pr}\left(Q_i^+>\epsilon M/K, x_i=0\right)=\mathbf{Pr}\left(Q_i^->\epsilon M/K, x_i=0\right)\\\notag
\leq& e^{-\epsilon M/K t}\left(1  + \frac{t(t-1)}{(2K-1)2!} +  \frac{t(t-1)(t-2)(t-3)}{(4K-3)4!}+...\right)^M\\\notag
=& \exp\left\{- \frac{M}{K}\left(\epsilon t  - K\log \left(1  + \frac{t(t-1)}{(2K-1)2!} +  \frac{t(t-1)(t-2)(t-3)}{(4K-3)4!}+...\right)\right)\right\}\\\notag
=&\exp\left\{- \frac{M}{K}H_1(t;\epsilon,K)\right\}
\end{align}
where
\begin{align}\notag
&H_1(t;\epsilon,K) = \epsilon t-K\log \left(1  + \frac{t(t-1)}{(2K-1)2!} +  \frac{t(t-1)(t-2)(t-3)}{(4K-3)4!}+...\right)\\\notag
&H_1(t;\epsilon,\infty) = \epsilon t  - \left(\frac{t(t-1)}{2\times 2!} +  \frac{t(t-1)(t-2)(t-3)}{4\times4!}+...\right)
\end{align}
Note that, by L'Hospital's Rule, we have
\begin{align}\notag
&\lim_{K\rightarrow\infty} \frac{\log \left(1  + \frac{t(t-1)}{(2K-1)2!} +  \frac{t(t-1)(t-2)(t-3)}{(4K-3)4!}+...\right)}{1/K}\\\notag
=&\lim_{K\rightarrow\infty} \frac{\frac{-2\frac{t(t-1)}{(2K-1)^22!}-4\frac{t(t-1)(t-2)(t-3)}{(4K-3)^24!}+...}{1  + \frac{t(t-1)}{(2K-1)2!} +  \frac{t(t-1)(t-2)(t-3)}{(4K-3)4!}+...}}{-1/K^2}
=\frac{t(t-1)}{2\times 2!} +  \frac{t(t-1)(t-2)(t-3)}{4\times4!}+...
\end{align}
This completes the proof.

\section{Proof of Lemma~\ref{lem_ProbH2}}\label{app_lem_ProbH2}

\begin{align}\notag
&\mathbf{Pr}\left(Q_i^+<\epsilon M/K, x_i>0\right)\\\notag
=&\mathbf{Pr}\left(\sum_{j=1}^M\log \left(1+sgn(y_j/s_{ij})\exp\left(-(K-1)w_{ij}\right)\right)<\epsilon M/K,x_i>0\right)\\\notag
=&\mathbf{Pr}\left(\exp\left(-t\sum_{j=1}^M\log \left(1+sgn(y_j/s_{ij})\exp\left(-(K-1)w_{ij}\right)\right)\right)>\exp\left(-t\epsilon M/K\right),x_i>0\right), \ \ t>0\\\notag
=&\mathbf{Pr}\left(\prod_{j=1}^M \left(1+sgn(y_j/s_{ij})\exp\left(-(K-1)w_{ij}\right)\right)^{-t}>\exp\left(-t\epsilon M/K\right),x_i>0\right)\\\notag
\leq& \exp\left(t\epsilon M/K\right)E^M\left( \left(1+sgn(y_j/s_{ij})\exp\left(-(K-1)w_{ij}\right)\right)^{-t};x_i>0\right)
\end{align}

Consider, for convenience,  $\alpha\rightarrow0$ and $x_i>0$. Again, we  study $sgn(y_j/s_{ij}) = sgn\left(x_i + \theta_i S_j/s_{ij}\right)$, where $S_j, s_{ij}\sim S(\alpha,1)$ i.i.d.  Let $T_{ij}  = sgn(y_j/s_{ij})\exp\left(-(K-1)w_{ij}\right)$. As $\alpha\rightarrow0$
\begin{align}\notag
T_{ij} =& sgn\left(x_i + \theta_i sgn(U_j)sgn(u_{ij})\left(\frac{w_{ij}}{W_j}\right)^{1/\alpha}\right) e^{-(K-1)w_{ij}}\\\notag
=& sgn\left(x_i + sgn(U_j)sgn(u_{ij})\left((K-1)\frac{w_{ij}}{W_j}\right)^{1/\alpha}\right) e^{-(K-1)w_{ij}}\\\notag
=&\left\{\begin{array}{ll}
sgn(x_i) e^{-(K-1)w_{ij}} &\text{ if } (K-1)w_{ij}<W_j \\
sgn(u_{ij}) e^{-(K-1)w_{ij}} &\text{ if } (K-1)w_{ij}>W_j
\end{array}
\right.
\end{align}
Thus,
\begin{align}\notag
&E\left( \left(1+sgn(y_j/s_{ij})\exp\left(-(K-1)w_{ij}\right)\right)^{-t};x_i>0\right)\\\notag
=&E\left\{ \int_{0}^{W_j/(K-1)}\left(1+\exp\left(-(K-1)u\right)\right)^{-t}e^{-u} du\right\}
+\frac{1}{2}E\left\{ \int_{W_j/(K-1)}^\infty\left(1+\exp\left(-(K-1)u\right)\right)^{-t}e^{-u} du\right\}\\\notag
+&\frac{1}{2}E\left\{ \int_{W_j/(K-1)}^\infty\left(1-\exp\left(-(K-1)u\right)\right)^{-t}e^{-u} du\right\}\\\notag
=&\frac{1}{2}\left\{ \int_{0}^{\infty}\left(1+\exp\left(-(K-1)u\right)\right)^{-t}e^{-u} du\right\}+\frac{1}{2}\left\{ \int_{0}^{\infty}\left(1-\exp\left(-(K-1)u\right)\right)^{-t}e^{-u} du\right\}\\\notag
+&\frac{1}{2}E\left\{ \int_0^{W_j/(K-1)}\left(1+\exp\left(-(K-1)u\right)\right)^{-t}e^{-u} du\right\}
-\frac{1}{2}E\left\{ \int_0^{W_j/(K-1)}\left(1-\exp\left(-(K-1)u\right)\right)^{-t}e^{-u} du\right\}\\\notag
=&\frac{1}{2}\int_{0}^{1}\left(1+u^b\right)^{-t} du+\frac{1}{2}\int_{0}^{1}\left(1-u^b\right)^{-t}e^{-u} du
-\frac{1}{2}\int_0^\infty e^{-w} \int_{w/b}^1\left[\left(1-u^b\right)^{-t} - \left(1+u^b\right)^{-t}\right] dudw
\end{align}
Again,  for convenience, we denote $b = K-1$.
\begin{align}\notag
& \int_0^1(1+u^b)^{-t} du
\\\notag
=&\int_0^1 1-u^bt + u^{2b}(-t)(-t-1)/2! + u^{3b}(-t)(-t-1)(-t-2)/3! + u^{4b}(-t)(-t-1)(-t-2)(-t-3)/4! + .... d u\\\notag
=&1 - \frac{t}{b+1} + \frac{t(t+1)}{(2b+1)2!} -  \frac{t(t+1)(t+2)}{(3b+1)3!}+\frac{t(t+1)(t+2)(t+3)}{(4b+1)4!}...
\end{align}
\begin{align}\notag
&\frac{1}{2} \int_0^1(1+u^b)^{-t} du+\frac{1}{2} \int_0^1(1-u^b)^{-t} du
=1  + \frac{t(t+1)}{(2b+1)2!} +\frac{t(t+1)(t+2)(t+3)}{(4b+1)4!}+...
\end{align}

For the other term, we have
\begin{align}\notag
&\frac{1}{2}\int_0^\infty e^{-w} \int_{w/b}^1\left[\left(1-u^b\right)^{-t} - \left(1+u^b\right)^{-t}\right] dudw\\\notag
=&\int_0^\infty e^{-w} \int_{e^{-w/b}}^1\left[t u^b +  t(t+1)(t+2)u^{3b}/3!+t(t+1)(t+2)(t+3)(t+4)u^{5b}/5!+..\right] dudw\\\notag
=&\left[\frac{t}{b+1}+\frac{t(t+1)(t+2)}{(3b+1)3!} + \frac{t(t+1)(t+2)(t+3)(t+4)}{(5b+1)5!}+...\right]\\\notag
-&\int_0^\infty e^{-w} \left[ \frac{t}{b+1}(e^{-w/b})^{b+1}+\frac{t(t+1)(t+2)}{(3b+1)3!}(e^{-w/b})^{3b+1} + \frac{t(t+1)(t+2)(t+3)(t+4)}{(5b+1)5!}(e^{-w/b})^{5b+1}+...\right] dw\\\notag
=&\left[\frac{t}{b+1}+\frac{t(t+1)(t+2)}{(3b+1)3!} + \frac{t(t+1)(t+2)(t+3)(t+4)}{(5b+1)5!}+...\right]\\\notag
-&\left[ \frac{t}{b+1}\frac{b}{2b+1}+\frac{t(t+1)(t+2)}{3!(3b+1)}\frac{b}{4b+1} + \frac{t(t+1)(t+2)(t+3)(t+4)}{5!(5b+1)}\frac{b}{6b+1}+...\right]
\end{align}
Combining the results yields
\begin{align}\notag
&E\left( \left(1+sgn(y_j/s_{ij})\exp\left(-(K-1)w_{ij}\right)\right)^{-t};x_i>0\right)\\\notag
=&1 - \frac{t}{b+1} + \frac{t(t+1)}{(2b+1)2!} -  \frac{t(t+1)(t+2)}{(3b+1)3!}+\frac{t(t+1)(t+2)(t+3)}{(4b+1)4!}-\frac{t(t+1)(t+2)(t+3)(t+4)}{(5b+1)5!}+...\\\notag
+&\left[ \frac{t}{b+1}\frac{b}{2b+1}+\frac{t(t+1)(t+2)}{3!(3b+1)}\frac{b}{4b+1} + \frac{t(t+1)(t+2)(t+3)(t+4)}{5!(5b+1)}\frac{b}{6b+1}+...\right]\\\notag
=&1 - \frac{t}{2b+1} + \frac{t(t+1)}{(2b+1)2!} -  \frac{t(t+1)(t+2)}{(4b+1)3!}+\frac{t(t+1)(t+2)(t+3)}{(4b+1)4!}-\frac{t(t+1)(t+2)(t+3)(t+4)}{(6b+1)5!}+...
\end{align}

Therefore, we can write
\begin{align}\notag
&\mathbf{Pr}\left(Q_i^+<\epsilon M/K, x_i>0\right)\leq \exp\left(-\frac{M}{K}H_2(t;\epsilon,K)\right)
\end{align}
where
\begin{align}\notag
&H_2(t;\epsilon,K) = -\epsilon t - K \log \left[1 + \sum_{n=2,4,6...}^\infty\frac{1}{n(K-1)+1} \prod_{l=0}^{n-1}\frac{t+l}{n-l}-
\sum_{n=1,3,5...}^\infty\frac{1}{(n+1)(K-1)+1} \prod_{l=0}^{n-1}\frac{t+l}{n-l}\right]\\\notag
&H_2(t;\epsilon,\infty) = -\epsilon t -  \left[\sum_{n=2,4,6...}^\infty\frac{1}{n} \prod_{l=0}^{n-1}\frac{t+l}{n-l}-
\sum_{n=1,3,5...}^\infty\frac{1}{(n+1)} \prod_{l=0}^{n-1}\frac{t+l}{n-l}\right]
\end{align}

\section{Proof of Lemma~\ref{lem_ProbH3}}\label{app_lem_ProbH3}
We introduce independent binary variables $r_j$, $j=1$ to $M$, so that $r_j = 1$ with probability $1-\gamma$. Define
\begin{align}\notag
Q_{i,\gamma}^{+} = \sum_{j=1}^M\log \left(1+{sgn(r_jy_j)}{sgn(u_{ij})}e^{-\left({K}-1\right)w_{ij}}\right)
= \sum_{j=1}^M\log \left(1+sgn\left({r_jy_j}/{s_{ij}}\right)e^{-\left({K}-1\right)w_{ij}}\right)
\end{align}
Note that $sgn(r_ju_{ij}) = 1$ with probability $1/2(1-\gamma)+1/2(\gamma)=1/2$, hence it has the same distribution as $sgn(u_{ij})$.  Following the proof of Lemma~\ref{lem_ProbH1}, we can derive
\begin{align}\notag
&\mathbf{Pr}\left(Q_{i,\gamma}^+>\epsilon M/K, x_i=0\right)\\\notag
=&\mathbf{Pr}\left(\sum_{j=1}^M\log \left(1+sgn(r_jy_j/s_{ij})\exp\left(-(K-1)w_{ij}\right)\right)>\epsilon M/K,x_i=0\right)\\\notag
=&\mathbf{Pr}\left(\sum_{j=1}^M\log \left(1+sgn(r_jS_j/s_{ij})\exp\left(-(K-1)w_{ij}\right)\right)>\epsilon M/K\right)\\\notag
=&\mathbf{Pr}\left(\sum_{j=1}^M\log \left(1+sgn(r_ju_{ij})\exp\left(-(K-1)w_{ij}\right)\right)>\epsilon M/K\right)\\\notag
=&\mathbf{Pr}\left(\prod_{j=1}^M \left(1+sgn(r_ju_{ij})\exp\left(-(K-1)w_{ij}\right)\right)>e^{\epsilon M/K}\right)\\\notag
=&\mathbf{Pr}\left(\prod_{j=1}^M \left(1+sgn(u_{ij})\exp\left(-(K-1)w_{ij}\right)\right)>e^{\epsilon M/K}\right)
\end{align}
At this point, it becomes the same as the problem in Lemma~\ref{lem_ProbH1}, hence we complete the proof.

\section{Proof of Lemma~\ref{lem_ProbH4}}\label{app_lem_ProbH4}

\begin{align}\notag
&\mathbf{Pr}\left(Q_{i,\gamma}^+<\epsilon M/K, x_i>0\right)\\\notag
=&\mathbf{Pr}\left(\prod_{j=1}^M \left(1+sgn(r_jy_j/s_{ij})\exp\left(-(K-1)w_{ij}\right)\right)^{-t}>\exp\left(-t\epsilon M/K\right),x_i>0\right)\\\notag
\leq& \exp\left(t\epsilon M/K\right)E^M\left( \left(1+sgn(r_jy_j/s_{ij})\exp\left(-(K-1)w_{ij}\right)\right)^{-t};x_i>0\right)
\end{align}
Consider $\alpha\rightarrow0$. We study $sgn(r_jy_j/s_{ij}) = sgn\left(x_ir_j + r_j\theta_i S_j/s_{ij}\right)$, where $S_j, s_{ij}\sim S(\alpha,1)$ i.i.d.  Let $T_{ij}  = sgn(r_jy_j/s_{ij})\exp\left(-(K-1)w_{ij}\right)$. As $\alpha\rightarrow0$
\begin{align}\notag
T_{ij} =& sgn\left(x_ir_j + r_j\theta_i sgn(U_j)sgn(u_{ij})\left(\frac{w_{ij}}{W_j}\right)^{1/\alpha}\right) e^{-(K-1)w_{ij}}\\\notag
=& sgn\left(x_ir_j + r_jsgn(U_j)sgn(u_{ij})\left((K-1)\frac{w_{ij}}{W_j}\right)^{1/\alpha}\right) e^{-(K-1)w_{ij}}\\\notag
=&\left\{\begin{array}{ll}
sgn(r_jx_i) e^{-(K-1)w_{ij}} &\text{ if } (K-1)w_{ij}<W_j \\
sgn(r_ju_{ij}) e^{-(K-1)w_{ij}} &\text{ if } (K-1)w_{ij}>W_j
\end{array}
\right.
\end{align}
Thus,
\begin{align}\notag
&E\left( \left(1+sgn(y_j/s_{ij})\exp\left(-(K-1)w_{ij}\right)\right)^{-t};x_i>0\right)\\\notag
=&(1-\gamma)E\left\{ \int_{0}^{W_j/(K-1)}\left(1+\exp\left(-(K-1)u\right)\right)^{-t}e^{-u} du\right\}
+\gamma E\left\{ \int_{0}^{W_j/(K-1)}\left(1-\exp\left(-(K-1)u\right)\right)^{-t}e^{-u} du\right\}\\\notag
&+\frac{1}{2}E\left\{ \int_{W_j/(K-1)}^\infty\left(1+\exp\left(-(K-1)u\right)\right)^{-t}e^{-u} du\right\}
+\frac{1}{2}E\left\{ \int_{W_j/(K-1)}^\infty\left(1-\exp\left(-(K-1)u\right)\right)^{-t}e^{-u} du\right\}\\\notag
=&\frac{1}{2}\left\{ \int_{0}^{\infty}\left(1+\exp\left(-(K-1)u\right)\right)^{-t}e^{-u} du\right\}+\frac{1}{2}\left\{ \int_{0}^{\infty}\left(1-\exp\left(-(K-1)u\right)\right)^{-t}e^{-u} du\right\}\\\notag
+&\left(\frac{1}{2}-\gamma\right)E\left\{ \int_0^{W_j/(K-1)}\left(1+\exp\left(-(K-1)u\right)\right)^{-t}e^{-u} du\right\}\\\notag
&-\left(\frac{1}{2}-\gamma\right)E\left\{ \int_0^{W_j/(K-1)}\left(1-\exp\left(-(K-1)u\right)\right)^{-t}e^{-u} du\right\}\\\notag
=&\frac{1}{2}\int_{0}^{1}\left(1+u^b\right)^{-t} du+\frac{1}{2}\int_{0}^{1}\left(1-u^b\right)^{-t}e^{-u} du
-\left(\frac{1}{2}-\gamma\right)\int_0^\infty e^{-w} \int_{w/b}^1\left[\left(1-u^b\right)^{-t} - \left(1+u^b\right)^{-t}\right] dudw
\end{align}
Again,  for convenience, we denote $b = K-1$. As shown in the proof of Lemma~\ref{lem_ProbH2}, we have
\begin{align}\notag
&\frac{1}{2} \int_0^1(1+u^b)^{-t} du+\frac{1}{2} \int_0^1(1-u^b)^{-t} du
=1  + \frac{t(t+1)}{(2b+1)2!} +\frac{t(t+1)(t+2)(t+3)}{(4b+1)4!}+...
\end{align}

For the other term, we have
\begin{align}\notag
&\int_0^\infty e^{-w} \int_{w/b}^1\left[\left(1-u^b\right)^{-t} - \left(1+u^b\right)^{-t}\right] dudw\\\notag
=&2\int_0^\infty e^{-w} \int_{e^{-w/b}}^1\left[t u^b +  t(t+1)(t+2)u^{3b}/3!+t(t+1)(t+2)(t+3)(t+4)u^{5b}/5!+..\right] dudw\\\notag
=&2\left[\frac{t}{b+1}+\frac{t(t+1)(t+2)}{(3b+1)3!} + \frac{t(t+1)(t+2)(t+3)(t+4)}{(5b+1)5!}+...\right]\\\notag
-&2\int_0^\infty e^{-w} \left[ \frac{t}{b+1}(e^{-w/b})^{b+1}+\frac{t(t+1)(t+2)}{(3b+1)3!}(e^{-w/b})^{3b+1} + \frac{t(t+1)(t+2)(t+3)(t+4)}{(5b+1)5!}(e^{-w/b})^{5b+1}+...\right] dw\\\notag
=&2\left[\frac{t}{b+1}+\frac{t(t+1)(t+2)}{(3b+1)3!} + \frac{t(t+1)(t+2)(t+3)(t+4)}{(5b+1)5!}+...\right]\\\notag
-&2\left[ \frac{t}{b+1}\frac{b}{2b+1}+\frac{t(t+1)(t+2)}{3!(3b+1)}\frac{b}{4b+1} + \frac{t(t+1)(t+2)(t+3)(t+4)}{5!(5b+1)}\frac{b}{6b+1}+...\right]\\\notag
=&2\left[\frac{t}{2b+1} + \frac{t(t+1)(t+2)}{3!(4b+1)}+\frac{t(t+1)(t+2)(t+3)(t+4)}{5!(6b+1)}+ ... \right]
\end{align}
Combining the results yields
\begin{align}\notag
&E\left( \left(1+sgn(y_j/s_{ij})\exp\left(-(K-1)w_{ij}\right)\right)^{-t};x_i>0\right)\\\notag
=&\left[1  + \frac{t(t+1)}{(2b+1)2!} +\frac{t(t+1)(t+2)(t+3)}{(4b+1)4!}+...\right]\\\notag
 &- \left(1-2\gamma\right)\left[\frac{t}{2b+1} + \frac{t(t+1)(t+2)}{3!(4b+1)}+\frac{t(t+1)(t+2)(t+3)(t+4)}{5!(6b+1)}+ ... \right]
\end{align}

Therefore, we can write
\begin{align}\notag
&\mathbf{Pr}\left(Q_{i,gamma}^+<\epsilon M/K, x_i>0\right)\leq \exp\left(-\frac{M}{K}H_4(t;\epsilon,K,\gamma)\right)
\end{align}
where
\begin{align}\notag
&H_4(t;\epsilon,K,\gamma) = -\epsilon t - K \log \left[1 + \sum_{n=2,4,6...}^\infty\frac{1}{n(K-1)+1} \prod_{l=0}^{n-1}\frac{t+l}{n-l}-
\sum_{n=1,3,5...}^\infty\frac{1-2\gamma}{(n+1)(K-1)+1} \prod_{l=0}^{n-1}\frac{t+l}{n-l}\right]\\\notag
&H_4(t;\epsilon,\infty,\gamma) = -\epsilon t -  \left[\sum_{n=2,4,6...}^\infty\frac{1}{n} \prod_{l=0}^{n-1}\frac{t+l}{n-l}-
\sum_{n=1,3,5...}^\infty\frac{1-2\gamma}{(n+1)} \prod_{l=0}^{n-1}\frac{t+l}{n-l}\right]
\end{align}

\newpage

{
%\bibliographystyle{abbrv}
%\bibliography{../bib/mybibfile}

\begin{thebibliography}{10}

\bibitem{Proc:Boufounos08}
P.~Boufounos and R.~Baraniuk.
\newblock 1-bit compressive sensing.
\newblock In {\em Information Sciences and Systems, 2008.}, pages 16--21, March 2008.

\bibitem{Article:Candes_Robust_JIT06}
E.~Cand\`{e}s, J.~Romberg, and T.~Tao.
\newblock Robust uncertainty principles: exact signal reconstruction from
  highly incomplete frequency information.
\newblock {\em IEEE Transactions on Information Theory}, 52(2):489--509, Feb
  2006.

\bibitem{Article:Chambers_JASA76}
J.~M. Chambers, C.~L. Mallows, and B.~W. Stuck.
\newblock A method for simulating stable random variables.
\newblock {\em Journal of the American Statistical Association},
  71(354):340--344, 1976.

\bibitem{Proc:Chen_AISTATS15}
S.~Chen and A.~Banerjee.
\newblock One-bit compressed sensing with the k-support norm.
\newblock In {\em AISTATS}, 2015.

\bibitem{Article:Chen98}
S.~S. Chen, D.~L. Donoho, Michael, and A.~Saunders.
\newblock Atomic decomposition by basis pursuit.
\newblock {\em SIAM Journal on Scientific Computing}, 20:33--61, 1998.

\bibitem{Article:Cressie_75}
N.~Cressie.
\newblock A note on the behaviour of the stable distributions for small index.
\newblock {\em Z. Wahrscheinlichkeitstheorie und Verw. Gebiete}, 31(1):61--64,
  1975.

\bibitem{Article:Donoho_CS_JIT06}
D.~L. Donoho.
\newblock Compressed sensing.
\newblock {\em IEEE Transactions on Information Theory}, 52(4):1289--1306,
  April 2006.

\bibitem{Proc:Frund_NIPS08}
Y.~Freund, S.~Dasgupta, M.~Kabra, and N.~Verma.
\newblock Learning the structure of manifolds using random projections.
\newblock In {\em NIPS}, Vancouver, BC, Canada, 2008.

\bibitem{Proc:1BitCS_ICML13}
S.~Gopi, P.~Netrapalli, P.~Jain, and A.~Nori.
\newblock One-bit compressed sensing: Provable support and vector recovery.
\newblock In {\em ICML}, 2013.

\bibitem{Article:Indyk_JACM06}
P.~Indyk.
\newblock Stable distributions, pseudorandom generators, embeddings, and data
  stream computation.
\newblock {\em Journal of ACM}, 53(3):307--323, 2006.

\bibitem{Article:1Bit_IT13}
L.~Jacques, J.~N. Laska, P.~T. Boufounos, and R.~G. Baraniuk.
\newblock Robust 1-bit compressive sensing via binary stable embeddings of
  sparse vectors.
\newblock {\em IEEE Transactions on Information Theory}, 59(4):2082--2102,
  2013.

\bibitem{Proc:Li_SODA08}
P.~Li.
\newblock Estimators and tail bounds for dimension reduction in $l_\alpha$
  ($0<\alpha\leq 2$) using stable random projections.
\newblock In {\em SODA}, pages 10 -- 19, San Francisco, CA, 2008.

\bibitem{Report:SymStableCode}
P.~Li.
\newblock Binary and multi-bit coding for stable random projections.
\newblock Technical report, arXiv:1503.06876, 2015.

\bibitem{Proc:Li_Hastie_NIPS07}
P.~Li and T.~J. Hastie.
\newblock A unified near-optimal estimator for dimension reduction in
  $l_\alpha$ $(0<\alpha\leq 2)$ using stable random projections.
\newblock In {\em NIPS}, Vancouver, BC, Canada, 2007.

\bibitem{Proc:CCCS_COLT14}
P.~Li, C.-H. Zhang, and T.~Zhang.
\newblock Compressed counting meets compressed sensing.
\newblock In {\em COLT}, 2014.

\bibitem{Article:Mallat93}
S.~Mallat and Z.~Zhang.
\newblock Matching pursuits with time-frequency dictionaries.
\newblock {\em IEEE Transactions on Signal Processing}, 41(12):3397 --3415,
  1993.

\bibitem{Article:Muthukrishnan_05}
S.~Muthukrishnan.
\newblock Data streams: Algorithms and applications.
\newblock {\em Foundations and Trends in Theoretical Computer Science},
  1:117--236, 2005.

\bibitem{Article:CoSaMP_09}
D.~Needell and J.~Tropp.
\newblock \text{CoSaMP}: Iterative signal recovery from incomplete and
  inaccurate samples.
\newblock {\em Applied and Computational Harmonic Analysis}, 26(3):301--321,
  2009.

\bibitem{Proc:Pati93}
Y.~Pati, R.~Rezaiifar, and P.~S. Krishnaprasad.
\newblock Orthogonal matching pursuit: recursive function approximation with
  applications to wavelet decomposition.
\newblock In {\em Signals, Systems and Computers, 1993. 1993 Conference Record
  of The Twenty-Seventh Asilomar Conference on}, pages 40--44 vol.1, Nov 1993.

\bibitem{Article:Plan_IT13}
Y.~Plan and R.~Vershynin.
\newblock Robust 1-bit compressed sensing and sparse logistic regression: A
  convex programming approach.
\newblock {\em IEEE Transactions on Information Theory}, 59(1):482--494, 2013.

\bibitem{Book:Samorodnitsky_94}
G.~Samorodnitsky and M.~S. Taqqu.
\newblock {\em Stable Non-Gaussian Random Processes}.
\newblock Chapman \& Hall, New York, 1994.

\bibitem{Proc:Slawski_NIPS15}
M.~Slawski and P.~Li.
\newblock b-bit marginal regression.
\newblock In {\em NIPS}, Montreal, CA, 2015.

\bibitem{Article:Zhang_RIP11}
T.~Zhang.
\newblock Sparse recovery with orthogonal matching pursuit under \text{RIP}.
\newblock {\em IEEE Transactions on Information Theory}, 57(9):6215 --6221,
  Sept. 2011.

\bibitem{Book:Zolotarev_86}
V.~M. Zolotarev.
\newblock {\em One-dimensional Stable Distributions}.
\newblock American Mathematical Society, Providence, RI, 1986.

\end{thebibliography}

}

\end{document}